\documentclass[traditabstract]{aa} % for the abstract without structuration
\usepackage{graphicx}
\usepackage{natbib}
\usepackage{epsfig}
\usepackage{txfonts}
\usepackage{longtable}
\def\ut#1{\mathop{\vtop{\ialign{##\crcr
     $\hfil\displaystyle{#1}\hfil$\crcr\noalign
     {\kern1pt\nointerlineskip}\hbox{$\hfil\sim\hfil$}\crcr
     \noalign{\kern1pt}}}}}

\def\undersymbol#1#2{\mathop{\vtop{\ialign{##\crcr
     $\hfil\displaystyle{#2}\hfil$\crcr\noalign
     {\kern1pt\nointerlineskip}\hbox{$\hfil#1\hfil$}\crcr
     \noalign{\kern1pt}}}}}

\def\degr{^0}

\begin{document}
   \title{An {\it XMM}-Newton search for X-ray sources in the Fornax dwarf galaxy}
   \subtitle{}
   \author{A.A. Nucita\inst{1,2}
          \and
          L. Manni\inst{1,2}
          \and
          F. De Paolis\inst{1,2}
          \and
          D. Vetrugno\inst{1,2}
          \and
          G. Ingrosso\inst{1,2}
          }
   \institute{Dipartimento di Matematica e Fisica ``Ennio De Giorgi'', Universit\`a del Salento, CP 193, I-73100 Lecce, Italy \\
              \and
              INFN, Sez. di Lecce, via Per Arnesano, CP 193, I-73100, Lecce, Italy
             %\email{}
             }
   \offprints{A.A. Nucita (\texttt{nucita@le.infn.it)}}
   \titlerunning{The {\it XMM}-Newton view of the Fornax dSph}
   \authorrunning{A.A. Nucita et al.}
   \date{Received  , ; accepted  , }

% \abstract{}{}{}{}{}
% 5 {} token are mandatory

\abstract{
We report the results of a deep archive {\it XMM}-Newton observation of the Fornax 
spheroidal galaxy that we analyzed with the aim of fully characterizing the X-ray source population 
(in most of the cases likely to be background active galactic nuclei)
detected towards the target. A cross correlation with the available databases allowed us to find 
a source that may be associated with a variable star belonging to the galaxy.
We also searched for X-ray sources in the vicinity of the Fornax globular clusters GC 3 and GC 4 
and found two sources probably associated with the respective clusters.
The deep X-ray observation was also suitable for the search of the intermediate-mass black hole 
(of mass $\simeq 10^{4}$ M$_{\odot}$) expected to
be hosted in the center of the galaxy. In the case of Fornax,
this search is extremely difficult since the galaxy centroid of gravity is poorly constrained because 
of the large asymmetry observed in the optical surface brightness.
Since we cannot firmly establish the existence of an X-ray counterpart of the putative black 
hole, we put constraints only on the accretion parameters. In particular, we found that the corresponding
upper limit on the accretion efficiency, with respect to the Eddington luminosity, is as low as a few $10^{-5}$.
}

   \maketitle
%
%________________________________________________________________

\section{Introduction}

Diffuse dwarf galaxies (DDGs) are low-luminosity galaxies
that seem to be characterized by
structural parameters (luminosity, stellar scale length)
fundamentally different from those found in
spiral and elliptical galaxies (\citealt{kormendy1985}) with the
dwarf spheroidal galaxies (dSphs) at the extreme end of this sequence.

In particular, dSphs are thought to be satellite galaxies in the Local Group
(see e.g., \citealt{mashchenko}), have approximately spheroidal shapes (sometimes typical of irregular
and late-type spiral galaxies), and are usually
at least two orders of magnitude less
luminous than the faintest known spiral galaxies. These systems have stellar contents in the range
$3\times10^3$ M$_{\odot}$ to $2\times10^7$ M$_{\odot}$ (\citealt{martin2008}) on
length scales of a few kpc or less. Additionally, they show evidence of
being dark-matter dominated at all radii (for a review see \citealt{mateo1997})
as shown by the measurements of the central velocity dispersion that are much larger than typical values 
(see \citealt{mateo1998a}, \citealt{kleina2001}, 
\citealt{kleina2002} and references therein). The central velocity dispersion allows the 
mass-to-light ratios to be estimated.

The differences between the normal galaxies and the dSphs families probably come from a different
formation history with the most favored theory being that dSphs have low mass density since past
supernova winds removed large amounts of gas (\citealt{silk1987}).
Despite being very different in their physical properties from spirals
and ellipticals, dSphs show kinematical properties that can be modeled using
dark matter (DM) halos with the same mass profiles as those which reproduce
the rotation curves of spirals (\citealt{salucci2012}). Thus, the derived central densities 
and core radii for dSphs are consistent with the values
obtained by extrapolation of the relevant quantities from spiral galaxies.

The dSph group is also interesting since it provides an optimal laboratory to study the evolution of a particular
stellar population (of known metallicity and age) without suffering extreme crowding conditions as often 
happens in globular clusters.
In this respect, the high-energy view of these galaxies, such as that offered by deep {\it XMM}-Newton observations, 
allows the study of the faint end of the
X-ray luminosity function of an old stellar population. Furthermore, by studying
the low mass X-ray binary 
(LMXB) population characteristics in dSphs and globular clusters, it is possible
to get information about the formation history (still challenging, see e.g.
\citealt{maccarone2005b}) of such systems. Since any persistently bright LMXB would entirely consume the mass of the
companion via accretion in a few hundred million years (see e.g. the discussion on the LMXB formation history 
in \citealt{maccarone2005b}), the presence of
bright X-ray binaries in old stellar systems represents a problem that is yet to be solved. An example of
these challenging targets is the Sculptor dSph galaxy. When studying a deep Chandra survey of this dwarf galaxy, \citet{maccarone2005b}
found at least five X-ray sources with optical counterparts hence pushing towards alternative formation theories of the local LMXB population.
A push in this direction would be the observation of targets with no globular cluster contamination (like the Sculptor galaxy) and, possibly,
with a short epoch of star formation.

Based on the extrapolation to globular clusters of the fundamental $M_{BH}-M_{Bulge}$ relation derived from the study of super massive
black holes in galactic nuclei (see e.g. \citealt{magorrian1998}), one expects to find intermediate 
mass black holes (hereafter IMBHs) in globular clusters,  i.e., spherical systems of stars
which survived the interactions with the surroundings objects and now orbit the center of the hosting galaxy. Since it is commonly
accepted that the galaxies and associated globular clusters
formed at the same time (see e.g. \citealt{globular1} and \citealt{globular2}), it is
natural to expect that at least some of these spherical systems may host an IMBH.

Apart from the DM and stellar population issues, dSphs are intriguing places to search for
IMBHs, i.e., collapsed objects in the mass range $10^2$-$10^5$ M$_{\odot}$
which are considered to be the missing link between the observed stellar mass black holes (of a few tens of solar masses) and the super massive ones
($10^6-10^8$ M$_{\odot}$) residing at the center of most galaxies. One of the reasons why the existence of such objects is expected is 
they might play a crucial role in the
formation of the super massive objects which are thought
to grow from a population of seed objects with masses in the IMBH range (see e.g. \citealt{ebisuzaki2001}).
Obviously, those seeds that did not accrete a
substantial amount of matter and/or did not merge to form a central super massive black hole remain as IMBHs.

We expect to find IMBHs in dSphs as well (\citealt{maccarone2005}). For example, \citet{reines2011} reported
that the nearby dwarf starburst galaxy Henize2-10 harbors a compact radio source at its dynamical 
center spatially coincident with a hard X-ray source (possibly an
$\simeq 10^6$ M$_{\odot}$ accreting black hole). \citet{farrell2009} found the brightest 
known ultra-luminous X-ray source HLX-1
(see e.g. \citealt{vandermarel2004} for a review) in the halo of the edge-on S0a galaxy ESO 243-49,
possibly associated with an IMBH whose mass was initially evaluated to be
$\ut> 9\times 10^3$ M$_{\odot}$ (\citealt{servillat2011}) and then better constrained to the range $9\times 10 ^3$-$9\times 10 ^4$ M$_{\odot}$
(\citealt{webb2012}). A further step towards the IMBH hypothesis was given by \citet{farrell2012} who detected evidence for a
young ($<$ 200 Myr) stellar cluster of total mass $\sim 10^6$ M$_{\odot}$ around the putative black hole and concluded that HLX-1
is likely to be the stripped remnant of a nucleated dwarf galaxy.

In the case of the Fornax dSph, \citet{volonteri} assumed that an IMBH of mass $M_{BH} \simeq 10^5$ M$_{\odot}$ is hosted in the galactic core
and suggested that measuring the dispersion velocity of the stars within $30$ pc from the center would
allow that hypothesis to be tested. \citet{jardel2012} recently constructed axisymmetric Schwarzschild
models in order to estimate the mass profile of the Fornax dSph and, once these models were tested versus the available kinematic data,
it was possible to put
a 1-$\sigma$ upper limit of $M_{BH} = 3.2\times 10^4$ M$_{\odot}$ on the IMBH mass.

In this work, we concentrate on the Fornax dSph re-analyzing a set of {\it XMM}-Newton data previously studied by \citet{orioproc} who searched
for the X-ray population in the Leo I and Fornax dSphs. In our work, we used the most recent analysis software and calibration files. 
Apart from the characterization of the high-energy population
detected towards the galaxy and the cross correlation with the available databases, we discuss the
possible identification of a few genuine X-ray sources belonging to the Fornax dSph.
We also considered the possible existence of an IMBH in the galaxy core as suggested by
 \citet{volonteri} and \citet{jardel2012} and show that one of the detected X-ray sources coincides with one of the possible Fornax dSph
centers of gravity. We then constrained the black hole accretion parameters and noted that additional important information may be obtained
by moderately deep radio observations and high-angular resolution X-ray data.

The paper is structured as follows: in Sect. 2 we describe the X-ray data analysis and in Sect. 3 we report our findings on the
 X-ray population observed towards the galaxy, its correlation with known catalogs,
and the identification of a few objects as genuine X-ray sources in Fornax dSph.
We further discuss the IMBH hypothesis and address our conclusions in Sect. 4.

\section{X-ray observations and data processing}

The Fornax dwarf galaxy (at J2000 coordinates RA = 02$^{\rm h}$ 39$^{\rm m}$ 59$\fs$3 and Dec = $-$34$^\circ$ 26$'$
$57.1''$) was observed on August 8, 2005 for $\simeq 100$ ks (Observation ID 0302500101)
with the three European Photon Imaging Cameras (EPIC MOS 1, MOS 2, and pn) (\citealt{struder2001}, \citealt{turner2001})
on board the {\it XMM}-Newton satellite. The target was observed in imaging mode with the full-frame
window and medium filter.

\subsection{Data reduction and screening}
The observation data files (ODFs) were processed using the {\it XMM}-Science
Analysis System (SAS version $11.0.0$\footnote{{\tt http://xmm.esa.int/sas/}}) with the
latest available calibration constituent files. The event lists for the three cameras were obtained
by processing the raw data via the standard {\it emchain} and {\it epchain} tools.

We followed standard procedures in screening the data. In particular, for the spectral analysis
we rejected time intervals affected by high levels of background activity. These time intervals
(particularly evident in the energy range 10--12 keV) were flagged, strictly
following the instructions described in the XRPS User's Manual\footnote{{\tt http://xmm.esac.esa.int/external/xmm\_user\_support/ \\
/documentation/rpsman/index.html}} by selecting a threshold of
0.4 counts s$^{-1}$ and 0.35 counts s$^{-1}$ for the pn and MOS cameras, respectively. 
This allowed us to compile lists of good time intervals (GTIs)
which were used to discard high background activity periods. The resulting exposure 
times for the two MOS and pn cameras were $\simeq 82$
ks and $\simeq 62$ ks, respectively. We screened the data by using the filter expressions 
${\rm \#XMMEA\_EM}$ (for MOS) and
$\#XMMEA\_EP$ (for pn). We also added the ${\rm FLAG==0}$ selection expression in order to 
reject events close to CCD gaps or bad pixels, taking into account
all the valid patterns (PATTERN in [0:12]) for the two MOS cameras and only single and 
double events (PATTERN in [0:4]) for pn.

\subsection{Source detection}

For each camera, the list of events was divided into 5 energy bands chosen according to
those used in the 2XMM catalog of serendipitous X-ray sources \citep{watson2009}, i.e., $B_1: 0.2-0.5$ keV, $B_2: 0.5-1.0$ keV,
$B_3: 1.0-2.0$ keV, $B_4: 2.0-4.5$ keV, and $B_5: 4.5-12.0$ keV. For the three EPIC cameras, we produced one image
for each energy band and a mosaic image in the $0.3-10$ keV energy band for inspection purposes only.

We then performed the source detection using the SAS task {\it edetect$\_$chain}. For each camera and input image, the tool first evaluates
the corresponding exposure map (via the task {\it eexpmap}) taking into account the calibration information
on the spatial quantum efficiency, filter transmission, and vignetting. In the next step, we produced image masks that delimited
the regions where the source searching was performed.

The sources identified with the local and map
searching algorithms\footnote{For more details, the reader is addressed to the on-line thread:\\ {\tt
 http://xmm.esac.esa.int/sas/current/documentation}} were then used by the task {\it emldetect}
which performs a point spread function (PSF) fitting in each of the EPIC cameras for the five energy bands simultaneously. In particular, for any of the detected sources, the location and extent were fixed to the same values in all
bands while leaving the count rates free to vary among the different energy bands.

After the task completion, three lists of sources remained (one per camera), each containing
the refined coordinates, count rates, hardness ratios, and maximum detection likelihood of the source candidates.
% % % % As usual, the hardness ratios and associated errors were evaluated respectively as
% % % % \begin{equation}
% % % % HR_i = \frac{B_{i+1}-B_{i}}{B_{i+1}+B_{i}}
% % % % \end{equation}
% % % % and
% % % % \begin{equation}
% % % % EHR_i = 2\frac{\sqrt{(B_{i+1}EB_i)^2+(B_{i}EB_{i+1})^2}}{(B_{i+1}+B_{i})^2}~,
% % % % \end{equation}
% % % % where $B_i$ and $EB_i$ are the count rates and associated errors in the energy band $i$ for $i=1$ to $4$.

For each of the detected sources, we derived the X-ray flux (in units of erg s$^{-1}$ cm$^{-2}$) in a given band as
\begin{equation}
F_i = \frac{B_{i}}{ECF_i}~,
\end{equation}
where $B_i$ is the count rate in the $i$ band and $ECF_i$ is an energy conversion factor
which has been calculated using the most recent calibration
matrices for the MOS 1, MOS 2, and pn. ECFs
for each camera, energy band, and filter are in units of $10^{11}$ counts cm$^{2}$ erg$^{-1}$.
In particular, we used the ECFs\footnote{For details on the adopted ECFs, see \citet{note2003}. Please note that,
since 2008, improvements in the calibration of EPIC cameras lead to changes in the ECFs. Thus,
the quoted ECFs must be multiplied by a correction factor to get a better agreement among the fluxes
evaluated in each camera \citep{note2006}.
The ECFs and associated correction factors are reported in the User's Guide of the 2XMM catalog of serendipitous sources available at\\
{\tt http://xmmssc-www.star.le.ac.uk/Catalogue/2XMMi-DR3}.}
obtained assuming a power-law model with photon index $\Gamma=1.7$ and
a Galactic foreground absorption of $N_H\simeq 3.0\times 10^{20}$ cm$^{-2}$ \citep{watson2009}. Note also that the adopted
hydrogen column density is of the same order of magnitude as the average value estimated towards the target
by using the ``N$_{\rm H}$" online
calculator\footnote{Available in the tool section of \tt http://heasarc.nasa.gov}, i.e.,
2.7$\times$10$^{20}$ cm$^{-2}$ \citep{kalberla2005}.

We refined the absolute astrometry by matching the candidate source lists to the USNO-B1 catalog \citep{usnob1}. From this catalog,
we extracted a sub-sample of stars with coordinates within 20$'$ from the Fornax center and used the SAS task {\it eposcorr} (with parameter of
maximum distance of 1$''$) to obtain the coordinate correction which was (on average) $-0.56'' \pm 0.50''$ in RA and $-1.77'' \pm 0.34''$ in Dec.
We purged the candidate source lists (one per EPIC camera) by accepting only sources with a maximum likelihood detection (as provided by the source detection
algorithm) larger than 10 (equivalent to 4$\sigma$) and then cross correlated the source lists
via the IDL routine {\it srcor.pro}\footnote{\tt{http://idlastro.gsfc.nasa.gov}} by requiring a critical radius of $1''$ outside which the correlations were rejected and
obtained the source parameters (i.e., coordinates, count rates, fluxes, and hardness ratios) by weighting (with the errors)
the values of interest associated with the sources identified in the three cameras.

After removing a few spurious sources (i.e., those positioned at the borders of the cameras and those not appearing as such in the 0.2-12 kev band), 
our catalog resulted in
107 sources detected towards the Fornax dSph galaxy with, in particular, 32 sources identified contemporarily in MOS 1, MOS 2, and pn,
4 sources only in MOS 1 and MOS 2, 7 sources only in MOS 1 and pn, 18 sources only in MOS 2 and pn, 30 sources only in pn, 
2 sources only in MOS 1, and 14 sources only in MOS 2. 
Note that the number of detected
sources agrees very closely with the 104 sources found by \citet{orioproc} when analyzing the same data set, the discrepancy probably 
arising from slightly different choices
in the data screening procedure and the detection threshold used.
%%%%%%%%%%%%%%%%%%%%%%%%%%%%%%%%%%%%%%%%%%%%%%%%%%%%%%%%%%%%%%%%%%%%%%%%%%
\begin{figure*}[!t]
\begin{center}
%\hspace{.2cm}
\psfig{figure=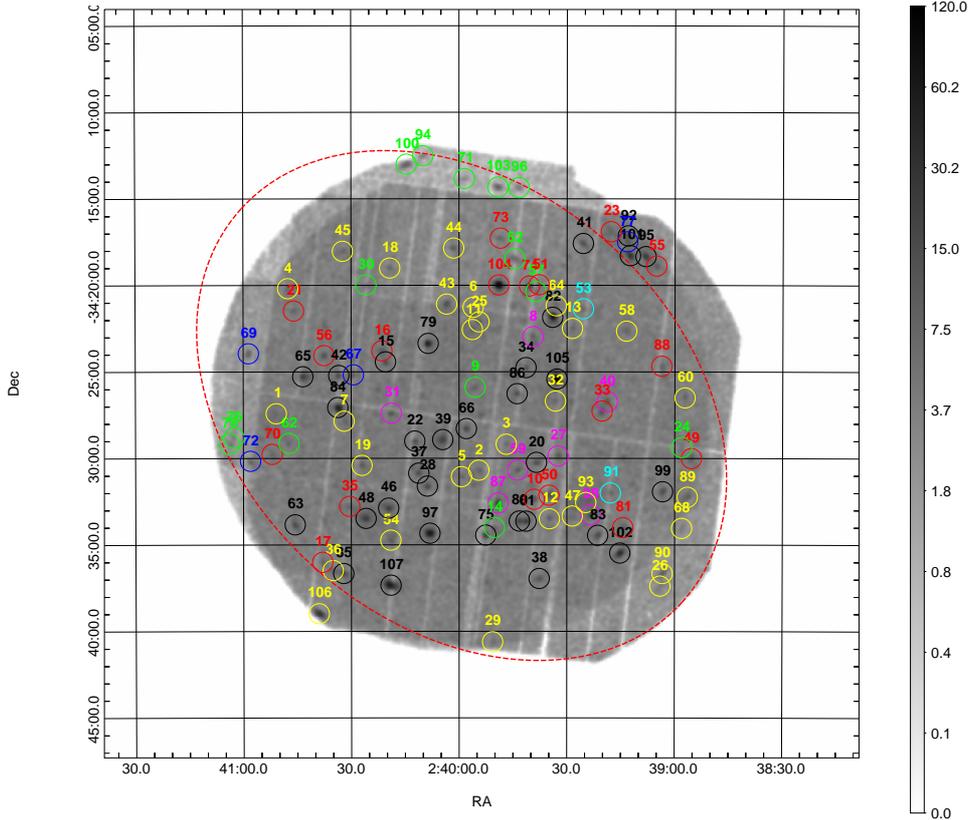,width=14.0cm,angle=0}
\end{center}
%\vspace{-0.7cm}
\caption{A mosaic image of the MOS 1, MOS 2, and pn exposures in the $0.3-10$ keV energy band (see text for details).}
\label{fig1}%
\end{figure*}
%%%%%%%%%%%%%%%%%%%%%%%%%%%%%%%%%%%%%%%%%%%%%%%%%%%%%%%%%%%%%%%%%%%%%%%%%%%
A mosaic image (logarithmically scaled and smoothed with a 3-pixel Gaussian kernel) of the MOS 1, MOS 2, and pn exposures
in the 0.3-10 keV energy band is given in Figure \ref{fig1}. Here, each of the
identified sources is indicated by a  $35''$ radius circle (containing $\cong$ 90\% of energy at 1.5 keV in the pn camera, see e.g. the
\citealt{xrps}), labeled with a sequential number and with a color
depending on which camera (or set of cameras) the source was detected by: yellow (pn), magenta (pn and MOS 1), red (pn and MOS 2), black (pn, MOS 1, and MOS 2),
blue (MOS 1 and MOS 2), cyan (MOS 1), and green (MOS 2). The red-dashed ellipse indicates the extension of the galaxy
which is characterized by semi-major and semi-minor axes of $\simeq 17'$ and $\simeq 13'$
(see the NASA/IPAC extragalactic database \footnote{$\tt http://ned.ipac.caltech.edu/$} -NED-) and a major-axis position angle of $\simeq 48\degr$ (\citealt{mateo1998b}).

In Table \ref{fornaxsources} we summarize our source analysis, ordering the list by increasing X-ray flux in the 0.2-12 keV energy band.
Here, we give a sequential number (Scr) for each source,
a label ($\#$) indicating which EPIC camera (or set of cameras) identified the source,
i.e., A (pn only), B (pn and MOS 1), C (pn and MOS 2), D (pn, MOS 1, and MOS 2), E (MOS 1 and MOS 2), F (MOS 1), and G (MOS 2).
We also give the J2000 coordinates with the associated errors (Column 3-5), the high-energy hardness ratios $HR_1$ and $HR_2$ (see next section) 
(Columns 6-7),
the $0.2-12$ keV absorbed flux (Column 8) and the cross correlations (Column 9-10).

We used the Two Micron All-Sky Survey (2MASS), the Two Micron All-Sky Survey Extended objects (2MASX) \citep{2mass},
the United States Naval Observatory all-sky survey (USNO-B1) \citep{usnob1}, and the variable star catalog in the Fornax galaxy \citep{varstar}
to correlate the X-ray source catalog with optical counterparts. In doing this, we associated with the coordinates of each of the 
identified X-ray source an error as resulting from the quadrature sum of the
{\it XMM}-Newton positional accuracy ($\simeq 2''$ at $2\sigma$ confidence level, see \citealt{kirsch2004}, and \citealt{guainazzi2010})
and the statistical error as determined by the {\it edetect$\_$chain} tool\footnote{Since the resulting
positional uncertainty is of a few arcseconds, we do not over-plot the source error circles in any of the figures appearing in the paper.}.
Similarly, the error associated with the optical counterpart was derived from the relevant catalog.

When an X-ray source is found to be within $1\sigma$ from an optical counterpart in a given catalog, we report in Table \ref{fornaxsources}
the corresponding distance in arcseconds in the relevant column. We remind the reader that the Fornax dSph galaxy was already the target
of a ROSAT X-ray observation (see \citealt{gizis1993} for details) with the purpose of characterizing the X-ray population (if any) in
the galaxy and of constraining the extended gas component.
% % % \footnote{By studying the X-ray surface brightness as a function of the
% % % distance form the Fornax dSph center, \citet{gizis1993} did not find any trend in the ROSAT data and put
% % % an upper limit flux of $\simeq 4\times 10^{-12}$ erg s$^{-1}$ cm$^{-2}$  (in the 0.5-2.0 keV band)
% % % from an extended gas component.}
The analysis of the ROSAT data resulted in the compilation of a catalog (hereinafter GMD093) listing 19
discrete sources congruous with the expected number of sources in the
extragalactic background. In Table \ref{tab} we give the GMD093 labels and coordinates (columns 1-3) of the sources
observed by ROSAT and the labels of the corresponding sources
in our catalog (column 4). The distance between the sources in the GMD093 catalog and the corresponding sources found in the
{\it XMM}-Newton list
is given in the fifth column with the error evaluated as the sum in quadrature between the {\it XMM}-Newton and ROSAT positional accuracies
\footnote{Because of the lack of any positional uncertainty on the sources
reported in \citet{gizis1993}, we associated each of the ROSAT detected
sources with an error of $\simeq 15''$ on both the celestial coordinates: we are aware
that this error may be underestimated since the positional accuracy of ROSAT PSPC
decreases with increasing offset from the detector axis.}. Finally, a possible classification
(based on the SIMBAD\footnote{$\tt http://simbad.u-strasbg.fr/simbad/$} and NED web sites)
is attempted in the sixth column (QSO -quasar-, EmG -emission line galaxy-, H2G -HII galaxy-, SyG -Seyfert galaxy-):
when a source is not cross correlated with SIMBAD and/or NED, this is simply labeled as {\it Xrs} (X-ray source).
Note that, among the 19 sources belonging to the GMD093 catalog, 6 are out of the {\it XMM}-Newton field of view.
Furthermore, from the $L_X$-$L_B$ relation ($L_B$ being the Fornax dSph B band luminosity), \citet{gizis1993}
estimated that at least one accreting binary should be present in the field of view.
For completeness, we mention that as far as the extended component is concerned, the ROSAT data did not show any diffuse
emission different from the normal X-ray background and no trend with radius was evident. Here, we concentrate on the point-like sources identified in the {\it XMM}-Newton observation of the Fornax dSph.
\addtocounter{table}{1}
\footnotesize{
\begin{center}
\begin{table*}
\begin{tabular}{|c|c|c|c|c|c|c|}
\hline
GMD093   & RA      &   Dec     & Src        &  Distance   & Type             & Ref.   \\
  ID     & (J2000) &  (J2000)  &            &  (arcsec)   &                  &        \\
\hline
1  &  2 40 15.7 & -34 13 23.1  & 100   & 27 $\pm $ 15      &   Xrs          & --       \\
2  &  2 39 14.7 & -34 15 04.8  &  --   &--               & Xrs             & --       \\
3  &  2 39 12.2 & -34 18 18.2  & 101   & 3 $\pm $ 15     &  Xrs           & --       \\
4  &  2 39 08.5 & -34 18 38.1  & 95    & 19 $\pm $ 15    &  Xrs           & --       \\
5  &  2 39 50.0 & -34 20 10.9  & 104   & 18 $\pm $ 15    &   QSO/EmG               & 1,2    \\
6  &  2 39 34.8 & -34 22 01.9  & 82    & 15 $\pm $ 15    &   Xrs          & --       \\
7  &  2 39 33.5 & -34 25 39.0  & 105   & 16 $\pm $ 15    &  QSO                    & 1,2    \\
8  &  2 40 34.2 & -34 27 11.0  & 84    & 8  $\pm $ 15    &  Xrs           & --       \\
9  &  2 38 20.8 & -34 30 17.6  & --    &--               & Xrs            & --       \\
10 &  2 39 26.2 & -34 32 46.7  & 93    & 23 $\pm $ 16   &   Xrs          & --      \\
11 &  2 40 08.1 & -34 34 25.3  & 97    & 6  $\pm $ 15    &  QSO                    & 2,3    \\
12 &  2 39 15.9 & -34 35 32.3  & 102   & 8 $\pm $  15    &   Xrs          & --       \\
13 &  2 39 03.7 & -34 36 46.4  & 90    & 8  $\pm $ 15    &   Xrs          & --       \\
14 &  2 38 48.8 & -34 36 59.9  & --    &--               &  Xrs           & --       \\
15 &  2 40 19.0 & -34 37 27.1  & 107   & 7  $\pm $ 15    &  QSO/H2G         & 1,2,4  \\
16 &  2 40 38.9 & -34 39 06.2  & 106   & 8  $\pm $ 15    &   QSO/SyG                   & 2,3    \\
17 &  2 42 08.1 & -34 40 06.8  & --    &--               & Xrs            & --       \\
18 &  2 38 55.5 & -34 40 52.4  & --    &--               & QSO                     & 1,2    \\
19 &  2 40 39.1 & -34 48 10.7  & --    &--               & Xrs             & --       \\
\hline
\end{tabular}
\caption {List of the detected X-ray sources cross-correlated with the catalog GMD093 (\citealt{gizis1993}). The GMD093 sources  labeled as 2,9,14,17,18, and 19
are out of the {\it XMM}-Newton FOV.}
\tablebib{
(1)~\citet{tinney1999}; (2) \citet{vcv2006}; (3) \citet{tdz1997}; (4) \citet{jones2006}.
\label{tab}}
\end{table*}
\end{center}
}

\section{The high-energy view of the Fornax dSph}

\subsection{X-ray colors and X-ray-to-NIR flux ratios}

\begin{figure}[!t]
\begin{center}
\hspace{-1.08cm}
\psfig{figure=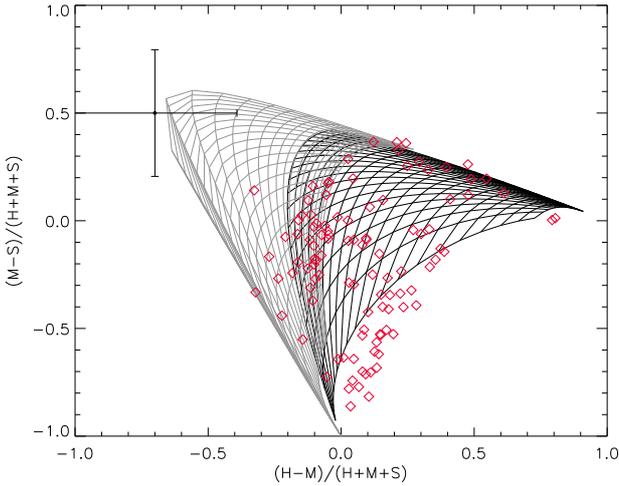,width=10.0cm,angle=0}
\end{center}
\vspace{-0.7cm}
\caption[]{The color-color diagram of the sources detected by the EPIC cameras towards the Fornax dSph galaxy. The solid lines represent the theoretical tracks
expected for different emitting models (see text for details). A representative error bar (obtained averaging all data point error bars) is also shown.}
\label{fig2}%
\end{figure}

Using the results presented in the previous section and with the purpose of a tentative classification of all the sources identified towards the Fornax dSph, we constructed the
color-color diagram in Figure \ref{fig2}. Here, we considered the colors
\begin{equation}
HR_1 = \frac{H-M}{S+M+H}~~{\rm and}~~HR_2 = \frac{M-S}{S+M+H}~,
\end{equation}
following the convention used in \citet{ramsay2006} in studying the Sagittarius and Carina dwarf galaxies and first
introduced by \citet{prestwich2003} and \citet{soria2003} S, M, and H correspond to the count rates in 0.3-1 keV,
1-4 keV, and 4-10 keV energy bands.

Since we expect that the detected X-ray sources are mainly accreting compact objects such as background AGN (Active Galactic Nuclei), X-ray binaries, and the brighter end of cataclysmic variables (CVs), we
compared the source measured hardness ratios (red squares) with two spectral models
(power-law, for simulating AGN and X-ray binaries, and bremsstrahlung, for CVs) which may be used to have an overall, although
simplified, description of the source spectra (see e.g. \citealt{ramsay2006} for a similar analysis made on the
Sgr and Car dSphs).
The model tracks, appearing in the color-color diagram, were obtained by simulating synthetic spectra within XSPEC 
(An X-Ray Spectral Fitting Package, \citealt{arnaud}), version 12.0.0. In particular, we give the
expected set of color-color contours for bremsstrahlung (grey
region) and power-law (black region) components. In both cases,
the equivalent hydrogen column density $N_H$ varies in the range
$10^{19}$cm$^{-2}$ to $10^{22}$ cm$^{-2}$: each of the almost horizontal
lines corresponds to models with equal $N_H$ which increases from bottom to top. The temperature kT of
the bremsstrahlung models (taken in the range  0.1 - 3 keV) and
the power-law index $\Gamma$ (in the range 0.1 - 3) is associated with
primarily vertical lines: the values of kT and $\Gamma$ increase from left to right and from
right to left, respectively.
A representative error bar, obtained by averaging all the
data point error bars, is also given. Most of the detected sources
have colors consistent with those of the absorbed power-law or
absorbed bremsstrahlung models, although a few of them may require combined spectra.
%%%%%%%%%%%%%%%%%%%%%%%%%%%%%%%%%%%%%%%%%%%%%%
%%%%%%%%%%%%%%%%%%%%%%%%%%%%%%%%%%%%%%%%%%%%%%
Most of the sources appear to have spectra consistent with that of a typical AGN (which in the hardness diagram of Figure \ref{fig2} would
correspond to a dot lying close to the harder color-color tracks, see e.g. \citealt{ramsay2006}).
However, because the large error bars affect the hardness ratio data, a classification based only on
the hardness ratio cannot constrain the nature of the objects in our sample.
% % % Most of the sources appear to have spectra consistent with that of a typical AGN (which in the hardness diagram of Figure \ref{fig2} would
% % % correspond to a dot lying close to the harder color-color tracks, see e.g. \citealt{ramsay2006})
% % % even if we cannot exclude a contamination of a few sources (X-ray binaries)
% % % hosted in the Fornax dSph galaxy.
% % % In fact, due to the large error bars affecting the hardness ratios, a classification based only on the hardness ratio cannot really
% % % constrain the nature of the objects in our sample and allow distinguishing among different types of objects such as X-ray binaries,
% % % CVs and stars.

\begin{figure}[!t]
\begin{center}
\hspace{-1.08cm}
\psfig{figure=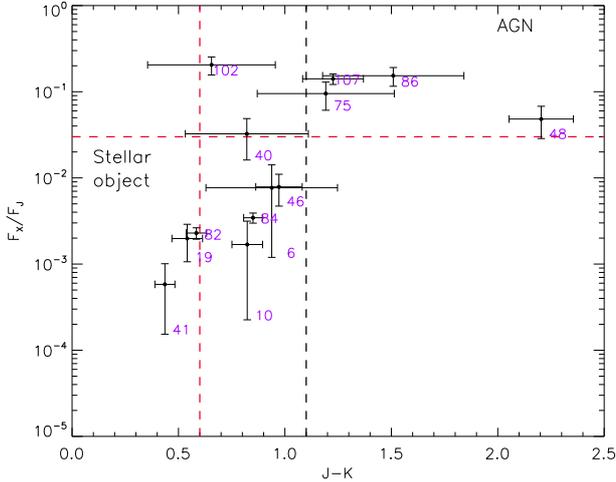,width=10.0cm,angle=0}
\end{center}
\vspace{-0.7cm}
\caption[]{The color-color diagram for the X-ray sources of our sample with a counterpart in the 2MASS catalog (see text for details).}
\label{figclass}%
\end{figure}

A better diagnostic tool was found by \citet{haakonsen2009} when studying the cross associations of the ROSAT All-Sky Survey Bright Source Catalog
(RASS/BSC) with the infrared counterparts found in 2MASS. In particular, these authors found that a color-color diagram 
(based on the ratio between the
$0.2-2.4$ keV flux ($F_X$) and the NIR flux in J band ($F_J$) versus the J-K color) is useful in studying the characteristics of the sources. 
In this plane, the galaxies
(Quasars and Seyfert 1 objects) and the coronally active stars (including pre-main sequence and main-sequence stars, high proper-motion objects,
and binary stars)
occupy distinct regions in such a way that galaxies have $(J-K)>0.6$ and $F_X/F_J>3\times 10^{-2}$ (dashed red lines in Figure \ref{figclass}), while almost all the objects in the second class have
$(J-K)<1.1$ (dotted black line in Figure \ref{figclass}) and $F_X/F_J<3\times 10^{-2}$, respectively\footnote{Note that
a better diagnostic would use the unabsorbed X-ray flux in the 0.2-2.4 keV band and the extinction-corrected NIR magnitudes: in the 
proposed color-color diagram of \citet{haakonsen2009},
the galaxies are positioned in the region described by $(J-K)>(J-K)_0>0.6$ and $(F_X/F_J)_0>F_X/F_J>3\times 10^{-2}$. On the contrary, the coronally active stars satisfy the relations
$(J-K)_0<(J-K)<1.1$ and $(F_X/F_J)<(F_X/F_J)_0<3\times 10^{-2}$, with the subscript {\it 0} indicating the result of the de-reddening procedure (see \citealt{haakonsen2009} for details).}.
For the sources of our sample which correlate with 2MASS
counterparts, we give in Table \ref{lasttable} the 0.2-2.4 keV band flux, the NIR J band flux and the J and K magnitudes. Note that the 0.2-2.4 keV band flux has been
obtained from that in the 0.2-12 keV band (given in Table 1) assuming in webPIMMS a power-law model with a spectral index $\Gamma=1.7$
and an absorption column density $N_H=3\,\times \,10^{20}\, {\rm cm^{-2}}$. In Figure \ref{figclass}, we give the color-color diagram 
(X-ray to J band flux ratio
against J-K) for the sources listed in Table \ref{lasttable}. The sources lying above the horizontal dashed line are consistent with background AGN,
while the others seem to have characteristics similar to X-ray active stars and binary sources: three of them (19, 41, and 82) have measured 
proper motions
in the PPMX catalog (see also Table 1), confirming the stellar nature of these objects. Note that the X-ray source labeled as 107 correlates 
(within 1.0$''$) with a source in the
2MASX and, according to our color-color diagram, is possibly associated with a background AGN (see also \citealt{mendez2011} where it is used as a reference background AGN
for proper motion estimates), while it was probably reported in the PPMX catalog by mistake 
(source labeled as 024019.0-343719): this confirms the predictive power of this color-color diagram.
\begin{table}
\footnotesize{
\begin{tabular}{|c|c|c|c|c|}
\hline
Src & F$_X$                    & F$_{J}$                 & J & K \\
    & ($\times 10^{-14}$ cgs)  & ($\times 10^{-13}$ cgs)   &   & \\
\hline
6  & 0.14 $\pm$     0.12  &     1.82 $\pm$     0.20  &     16.12 $\pm$   0.11 &     15.18 $\pm$   0.20 \\
10 & 0.18 $\pm$     0.16  &     10.90 $\pm$    0.32  &     14.17 $\pm$   0.03 &     13.35 $\pm$   0.04\\
19 & 0.34 $\pm$     0.16  &     17.11 $\pm$    0.43  &     13.68 $\pm$   0.03 &     13.14 $\pm$   0.04\\
40 & 0.57 $\pm$     0.28  &     1.77 $\pm$     0.16  &     16.15 $\pm$   0.10 &     15.32 $\pm$   0.19\\
41 & 0.59 $\pm$     0.45  &     102.71 $\pm$   2.30  &     11.74 $\pm$   0.02 &     11.30 $\pm$   0.02 \\
46 & 0.69 $\pm$     0.28  &     8.86 $\pm$     0.43  &     14.40 $\pm$   0.05 &     13.42 $\pm$   0.10 \\
48 & 0.74 $\pm$     0.28  &     1.53 $\pm$     0.20  &     16.30 $\pm$   0.15 &     14.10 $\pm$   0.15 \\
75 & 1.42 $\pm$     0.47  &     1.50 $\pm$     0.21  &     16.33 $\pm$   0.15 &     15.14 $\pm$   0.18\\
82 & 1.75 $\pm$     0.26  &     76.42 $\pm$    1.70  &     12.06 $\pm$   0.02 &     11.47 $\pm$   0.02\\
84 & 1.87 $\pm$     0.25  &     54.25 $\pm$    1.10  &     12.43 $\pm$   0.02 &     11.58 $\pm$   0.02 \\
86 & 2.03 $\pm$     0.38  &     1.32 $\pm$     0.22  &     16.46 $\pm$   0.17 &     14.95 $\pm$   0.20 \\
102& 4.25 $\pm$     0.93  &     2.10 $\pm$     0.19  &     15.97 $\pm$   0.10 &     15.32 $\pm$   0.20 \\
107& 9.84 $\pm$     1.22  &     6.97 $\pm$     0.48  &     14.66 $\pm$   0.10 &     13.43 $\pm$   0.10 \\
\hline
\end{tabular}
\caption{A few sources of the X-ray sample detected towards the Fornax dSph correlate with the 2MASS catalog. Following
\citet{haakonsen2009}, we can try to constrain the nature of the sources by using the X-ray (in the 0.2-2.4 keV band) and NIR (J and K bands) fluxes
(see text and Figure \ref{figclass} for details).
}
\label{lasttable}
}
\end{table}
%%%%%%%%%%%%%%%%%%%%%%%%%%%%%%%%%%%%%%%%%%%%%%

\subsection{Background sources from the log N - log S plot}
We estimated the number of background sources expected towards the Fornax dwarf galaxy through the log N - log S diagram \citep{Hasinger} 
and by using
the minimum absorbed flux in the 0.2-12 keV band among the sources of our analysis, i.e., 
$F^{Abs}_{0.2-12}=2.24\,\times\,10^{-15} {\rm erg\, s^{-1}\, cm^{-2}}$.
Using webPIMMS v3.9\footnote{{\tt http://heasarc.gsfc.nasa.gov/Tools/w3pimms.html}} and assuming a power-law model with a spectral index $\Gamma=1.7$
and an absorption column density $N_H=3\,\times \,10^{20}\, {\rm cm^{-2}}$, we obtained an unabsorbed flux in
the same band $F^{Unabs}_{0.2-12}=2.47\,\times\,10^{-15} {\rm erg\, s^{-1}\, cm^{-2}}$,
hence $F^{Unabs}_{0.5-2}=6.95\,\times\,10^{-16} {\rm erg\, s^{-1}\, cm^{-2}}$ in the 0.5-2 keV band.

We used this value as an input parameter of the \citet{Hasinger} method, obtaining a theoretical number of AGN as a function of the angular distance from the galactic center. Then we
divided the Fornax field of view (FOV) into five rings and compared the number of sources observed in each annulus to the expected number using the
log N - log S
relation \citep{Hasinger}, see Table \ref{tab2}. The number of sources detected in each annulus is consistent with the expected
number except in the external annulus, as the expected value does not account for border effects although all the detected sources in the Fornax FOV may be background objects, because of the intrinsic 
statistical meaning of the
log N - log S diagram,
it cannot be ruled out that some of them actually belong to the galaxy itself. This conclusion is somehow supported by Figure \ref{figclass} where the X-ray sources
in the bottom-left region are most likely of local origin.

\subsection{Hint of a local variable source associated with an X-ray source}

When searching for optical counterparts to the detected X-ray sources (see Table \ref{fornaxsources}) in
the available catalogs, we found that source number 61 is possibly associated, within 2$"$,
with one source (J023941.4-343340) belonging to a catalog of variable stars \citep{varstar}, making it a good candidate for a genuine
X-ray source in the Fornax galaxy. The angular sensitivity of {\it XMM}-Newton, however, makes it difficult to separate this source from the one labeled 80 (see Figure \ref{fig1}) which is only $\simeq 22"$ apart.
One solution is to reduce the contamination from the nearby source by reducing the source extraction radius and by selecting suitable
background regions. In the particular case of source 61, we used a source extraction circle
with a radius of $\simeq 11"$ (corresponding to $\simeq 50\%$ of the source encircled energy, see \citealt{xrps}) and extracted
the background from a circular region having the same radius as the source extraction area
and localized at a distance of $\simeq 22"$ from source 80. The latter choice allowed us to properly
remove the background noise as well as any pattern hidden in the data caused by the proximity of source 80.
We then extracted the pn spectrum of source 61 (which consisted of only $\simeq 200$ counts)
and fitted it with an absorbed power-law model with hydrogen column density fixed to the value 2.7$\times$10$^{20}$ cm$^{-2}$
(see Section 2.2). The best fit, which converged to a power-law index $\Gamma \simeq 2.0$, corresponded to a flux in the $0.2-12$ keV energy
band of $\simeq 2.0 \times 10^{-14}$ erg s$^{-1}$ cm$^{-2}$ consistent with that obtained
from the analysis discussed in Section 2 (see also Table 1).

Using the extraction regions previously described, we generated source and background pn light curves (with a bin-size of $500$ seconds)
from the cleaned event files and used the SAS task {\it epiclccorr} to account for absolute
and relative corrections and for the background subtraction. The resulting light curve was then
analyzed with the Lomb-Scargle method (\citealt{scargle1982}) to search for periodicities in the range $1000-10000$ seconds and
the significance of any possible features appearing in the Lomb-Scargle periodogram was evaluated by
comparing the peak height with the power threshold corresponding to a given false alarm probability in white noise simulations
\citep{scargle1982}. When applied to the presently available data, this method did not detect any clear periodicity.
% % % A better characterization - including a detailed time analysis of the high-energy data -
% % % would be achieved using, for example, the better angular resolution of the Chandra observatory.

\begin{table}
\footnotesize{
\begin{tabular}{|c|c|c|c|c|}
\hline
Annulus & R$_{in}$ & R$_{ex}$ & \# Exp & \# Obs       \\
        & (arcmin) & (arcmin) &        &              \\
\hline
  1 & 0.00 &  0.76  &  0.4 $\pm$ 0.1  &  0 \\
  2 & 0.76 &  3.60  & 10.2 $\pm$ 1.7  &  6 \\
  3 & 3.60 &  6.50  & 24.2 $\pm$ 4.0  & 21 \\
  4 & 6.50 &  9.00  & 32.1 $\pm$ 5.0  & 29 \\
  5 & 9.00 & 16.00  & 144.5$\pm$ 20.0 & 51 \\
\hline
\end{tabular}
\caption {List of sources expected through the log N - log S diagram and observed in annuli around Fornax center. Here,
$R_{in}$ and $R_{ex}$ represent the interior and exterior annulus radii, respectively.}
 \label{tab2}}
\end{table}

\subsection{The Fornax globular clusters}
\begin{figure}[ht]
\centering
\includegraphics[width=100mm]{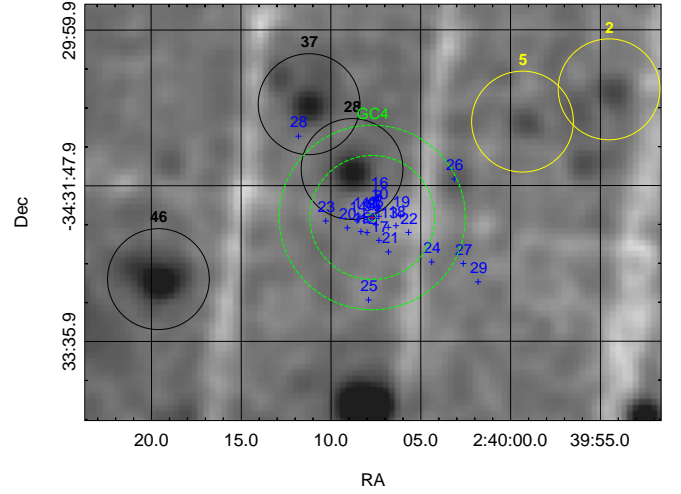}
\caption{A zoomed view of the region around the Fornax globular cluster GC 4 (see text for details).}
\label{figGC4}
\end{figure}

\begin{figure}[ht]
\centering
\includegraphics[width=100mm]{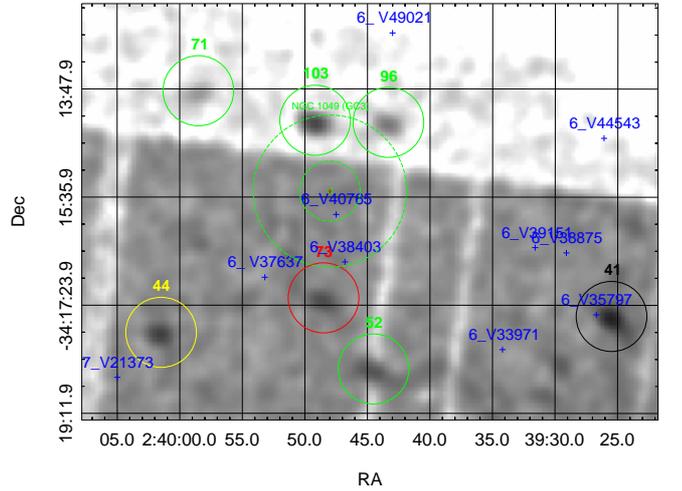}
\caption{As in Figure \ref{figGC4}, but the circles are centered on the coordinates of the globular cluster GC 3.}
\label{figGC3}
\end{figure}

Among all the other dSphs which are satellites of the Milky Way, the Fornax dSph has the peculiar characteristic of
hosting five globular clusters (GCs) whose
(J2000) coordinates and main structural parameters
(mass $M$, core radius $r_c$, projected distance $R_p$ from the galaxy center, tidal radius $r_t$, 
and maximum radius $r_M$) are reported in Table \ref{tableGC}.
Apart from the puzzling question of why dynamical friction has not yet
dragged any of the Fornax GCs towards the center of the galaxy (see e.g. \citet{cole2012} for an introduction to this intriguing problem and
to \citealt{goerdt2006} for a possible solution\footnote{\citet{goerdt2006} have shown that dynamical friction cannot drag globular clusters
to the galaxy center in the case of cored dark-matter halo distributions. In this case the drag is stopped at the point where the dark-matter
density remains constant, i.e., $\simeq 200$ pc.}), a search for X-ray sources associated with the GCs is also interesting.
We performed this search towards the five Fornax dSph GSs and found that a few X-ray sources
are located close to GC 3 and GC 4 (see the zoomed views of the {\it XMM}-Newton field around
the two clusters in Figures \ref{figGC4} and \ref{figGC3}.

For GC 4, the three green dashed concentric circles are centered on the GC coordinates and have radii of $2.6''$, $43''$, and $64''$, corresponding to
the cluster core radius, tidal radius, and maximum radius (see also Table \ref{tableGC}) as given in \citet{cole2012} and \citet{mackey}. In the case of
GC 3, for clarity, we have plotted only the circles having a radius equal to the core radius ($2.4''$) and tidal radius ($77''$).
\begin{table}
\footnotesize{
\begin{tabular}{|c|c|c|c|c|c|c|c|}
\hline
GC  & RA       & Dec   & M                    & $r_c$& $R_p$ & $r_T$ &  $r_M$         \\
    & (J2000)  & (J2000)&                     & (pc) & (kpc) & (pc)  & (pc)          \\
\hline
% 1& 2 37 02    & -34 11 00 & 0.37& 10.03& 1.69 & {37.12} & {49.71} \\
% 2& 2 38 44   & -34 48 36 & 1.82& 5.81& 1.05   & {49.05} & {50.38} \\
% 3& 2 39 48   & -34 15 24 & 3.63& 1.60& 0.43   & {51.04} & {50.38} \\
% 4& 2 40 7.7  & -34 32 10 & 1.32& 1.75& 0.24   & {28.50} & {42.42} \\
% 5& 2 42 21   & -34 06 12 & 1.78& 1.38& 1.43   & {49.05} & {50.38} \\
1& 2 37 02    & -34 11 01 & 0.4& 10.0& 1.7 & {37.1} & {49.7} \\
2& 2 38 44   & -34 48 30 & 1.8& 5.8& 1.0   & {49.0} & {50.4} \\
3& 2 39 48   & -34 15 30 & 3.6& 1.6& 0.4   & {51.0} & {50.4} \\
4& 2 40 7.7  & -34 32 11 & 1.3& 1.7& 0.2   & {28.5} & {42.4} \\
5& 2 42 21   & -34 06 06 & 1.8& 1.4& 1.4   & {49.0} & {50.4} \\
\hline
\end{tabular}
\caption {The coordinates of the Fornax globular clusters are taken from the NASA/IPAC extragalactic database (NED). For each cluster we give
the mass $M$ (in units of $10^5$ M$_{\odot}$), core radius $r_c$, and projected distance $R_p$ from the galaxy center as found in \citet{cole2012},
while the values of the tidal radius $r_t$ and maximum radius $r_M$ (expressed in parsec for a distance of $0.138$ Mpc) were derived from \citet{mackey} and references therein.}
\label{tableGC}
}
\end{table}
All the other circles (with associated radii of $35''$) appearing in the figures indicate the X-ray sources detected in our analysis (see Sect. 2.2).

In the case of GC 4, we identified one X-ray source (labeled as 28 in the source list appearing in Table \ref{fornaxsources}) which
is at a distance of $\simeq 37''$ from the globular cluster center\footnote{ We checked that none of the X-ray sources of our sample
has a counterpart in the catalog of variable stars (indicated as blue crosses in Figure \ref{figGC4}) as reported by \citet{greco2007}.}. Consequently, considering the GC 4 structural parameters reported in
Table \ref{tableGC}, one observes that source 28 is well within the tidal radius ($\simeq 43''$) of the globular cluster and possibly associated with it, as already noted
by \citet{orioproc} when analyzing the same {\it XMM}-Newton data set.

In the case of GC 3, two variable Sx Phe sources\footnote{A Sx Phe object is a variable pulsating star whose magnitude
can vary from a few $0.001$ mag to several $0.1$ mag with typical periods of $P\ut < 0.10$ days and is often used as a distance estimator. For a description on the main properties
of this class of variable stars we refer the reader to \citet{pych} and references therein.} (6$\_$V40765 and 6$\_$V38403)
were identified by \citet{poretti} as belonging to the globular cluster. Both sources appear to be at a distance of $\simeq 30''$ and $\simeq 78''$ from the globular cluster center and, therefore,
are possibly associated with GC 3.  As one can see from a close inspection of Figure \ref{figGC3}, 
there is no clear association of any of the variable stars (blue crosses) 
observed in GC 3 with
the detected X-ray sources. Moreover, one X-ray source (labeled as 103 at $\simeq 72''$ 
from the globular cluster center) and identified in the {\it XMM}-Newton field of view is within
the globular cluster tidal radius ($\simeq 77''$) and, therefore, may be associated with it (as already claimed by \citealt{orioproc}).
%
% Here, we also note that source 103 was already associated with GC 3 by \citealt{orioproc} although the estimated distance from the globular cluster  center
% was $\simeq 88''$: the difference being probably due to different globular cluster center coordinates used in the analyses.
%
Finally, we note that for an estimated distance of
0.138 Mpc to the Fornax dSph, the observed 0.2-12 keV band luminosities associated with the sources 28 and 103 are
$\simeq 2.3 \times 10^{34}$ erg s$^{-1}$ and $\simeq 2.6 \times 10^{35}$ erg s$^{-1}$, respectively. While the source possibly associated with GC 4 has a luminosity
comparable to that of a typical CV, the source possibly residing in GC3 has a luminosity well within the typical ranges 
(although towards the lower limit)
for LMXBs and high-mass X-ray binaries (HMXBs) (see
e.g. \citealt{fabbiano}, \citealt{kuulkers}, and \citealt{tauris}).
% % %
% % % As a last remark, we note that in the catalog from \citet{poretti} we found a number of variable star targets which are close
% % % (within a few arcseconds) to some X-ray sources found with the present analysis.
% % % Although the correlation could be only by chance, we give in Table \ref{tableporetti} the list of the possible associations without a further discussion.
% % % Here, the first and second columns represent the label given to the X-ray source (in accordance with Table
% % % \ref{fornaxsources}) and the label given to the variable star. In the third column, we also give the distance (in arcseconds) between the
% % % corresponding source centroids.
% % %
% \begin{figure}[htbp]
% \vspace{5.0cm} \special{psfile=fig3.eps vscale=70. hscale=70. voffset=-200 hoffset=-70.0 angle= 0.0}
% \caption{A zoomed view of the region around the Fornax globular cluster GC 4 (see text for details).}
% \label{figGC4}
% \end{figure}
%
%%%%%%%%%%%%%%%%%%%%%%%%%%%%%%%%%%%%%%%%%%%%%%
%%%%%%%%%%%%%%%%%%%%%%%%%%%%%%%%%%%%%%%%%%%%%%
%%%%%%%%%%%%%%%%%%%%%%%%%%%%%%%%%%%%%%%%%%%%%%

\subsection{An intermediate-mass black hole in the Fornax galaxy: high-energy constraints}

When the galactic globular clusters were identified in X-rays,
\citet{bahcall1975} and \citet{silk1976} suggested that the observed emission was due to the
presence of IMBHs in the mass range $\sim 10^2$ M$_{\odot}$ -- $10^5$M$_{\odot}$ accreting material from the
intracluster medium.

The discovery of ultra-luminous compact X-ray sources (ULXs, with
luminosity greater than $\sim 10^{39}$ erg s$^{-1}$) initially pushed the community to interpret such objects
as IMBHs. Note however that the current accepted sceario is that all the ULXs (with the exception of the highest luminosity objects still
contain an IMBH) are stellar-sized black holes which accrete at super-Eddington rates (\citealt{miller2003}).
More evidence comes from the study of the central velocity dispersion of stars in specific globular clusters 
(as G1 in the Andromeda galaxy, see e.g. Gebhardt et al. \citeyear{gebhardt2002}, but also Pooley et al. \citeyear{pooley2006},
or M15 and $\omega$ Centauri in the Milky Way, see e.g. Gerssen et al. \citeyear{gerssen2002}, \citeyear{gerssen2003}, and  
Miocchi \citeyear{miocchiomegacen}, respectively ) which may contain
central IMBHs. As far as M15 is concerned, milli-second pulsar timing studies have allowed us 
to put an upper limit of $\simeq 3\times 10 ^{3}$ M$_{\odot}$ on the IMBH mass
(\citealt{depaolis1996}). Finally, compact objects of IMBH size are also predicted by N-body simulations 
(see e.g. \citealt{potegiezwart2004}) as
a consequence of merging of massive stars.
% % % %%%%%%%%%%%%%%%%%%%%%%%%%%%%%%%%%%%%%%%%%%%%%%
% % % \begin{table}
% % % \footnotesize{
% % % \begin{tabular}{|c|c|c|}
% % % \hline
% % % Src  & Sx Phe stars & Distance        \\
% % %      &              & (arcsec)        \\
% % % \hline
% % % {34}    &$6\_V10166$      &    9.0             \\
% % % {43}    &$6\_V35797$      &   16.0             \\
% % % {65}    &$8\_V4532$       &   25.0            \\
% % % {67}    &$7\_V6056$       &   11.0            \\
% % % {105}    &$6\_V8812$      &   24.0            \\
% % % \hline
% % % \end{tabular}
% % % \caption{A few Sx Phe stars in the catalog from \citet{poretti} are found within a few arcseconds from an
% % % X-ray source in the field of view.}
% % % \label{tableporetti}
% % % }
% % % \end{table}
As noted by several authors (see e.g. \citealt{baumgardt2005}, and \citealt{miocchi}), photometric studies may provide a further 
hint for the existence of IMBHs in globular clusters. 
It is expected
that the mass density profile of a stellar system with a central IMBH follows a cuspy $\rho \propto
r^{-7/4}$ law, so that the projected density profile, as well as the
surface brightness, should also have a cusp profile with slope $-3/4$.

As shown by Miocchi (\citeyear{miocchi}), the globular
clusters that most likely host a central IMBH are those characterized by a projected photometry well fitted by a King profile, except in the
central part where a power-law deviation ($\alpha\simeq -0.2$) from a flat behavior is expected. However, the errors with which the slopes
of central densities can be determined are approximately $0.1-0.2$ (\citealt{noyola}), so that optical surface density profiles do not give
clear evidence of the existence of IMBH in globular clusters.

It is then natural to expect that observations in different bands of the electromagnetic spectrum, such as radio and X-ray bands, would 
permit further constraints on the IMBH parameters. This issue was considered by several authors. For example, \citet{grindlay2001} provided the census of the compact objects
and binary populations in the globular cluster 47 Tuc and obtained an upper limit
to the central IMBH of a few hundred solar masses; \citet{nucita2008} showed that
the core of the globular cluster NGC 6388 harbors several {\it X}-ray sources;
and \citet{cseh2010} refined the analysis putting an upper limit to the central IMBH mass of a few thousand solar masses
(see also \citealt{bozzo2011}, and \citealt{nucita2012}).

One expects to find IMBHs in dSph as well (\citealt{maccarone2005}).
In the specific case of the Fornax dSph, \citet{volonteri} assumed that an IMBH of mass $M_{BH} \simeq 10^5$ M$_{\odot}$ exists in the galactic core
and suggested that measuring the dispersion velocity of the stars within $30$ pc from the center would allow us to test that hypothesis.
Furthermore, \citet{jardel2012} recently constructed axisymmetric Schwarzschild
models in order to estimate the mass profile of the Fornax dSph. These models were tested versus the available kinematic data allowing the authors to put
a 1-$\sigma$ upper limit of $M_{BH} = 3.2\times 10^4$ M$_{\odot}$ on the IMBH mass. We will use the latter
value in the following discussions.

\begin{figure}[ht]
\centering
\includegraphics[width=100mm]{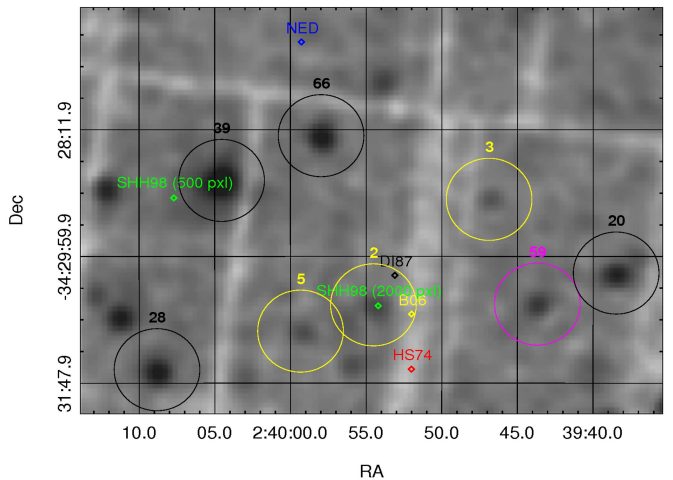}
\caption{A zoomed view around the Fornax dSph center. The diamonds indicate the global centroids of the galaxy obtained by several authors. In particular,
the green diamonds (with labels {\it SHS98 500 pxl} and {\it SHS98 2000 pxl}) correspond to the galaxy centers obtained by \citet{stetson}, the red and black diamonds
indicate the centroids as derived by \citet{hodge1974} and \citet{demers}, respectively (see text for details).}
\label{figcenters}
\end{figure}

Since any Brownian motion of IMBH at the center of the galaxy is negligible\footnote{
From simulations, one expects an IMBH within a globular cluster (or a spheroidal galaxy) to move randomly when interacting with the surrounding stars.
In the assumption that all the stars have the same mass $m$, the IMBH
moves with an amplitude $\sim r_c(m/M_{BH})$ (see e.g. \citealt{bw}, \citealt{gurzadyan}, and \citealt{merritt}) where $r_c$ is the core radius
and $M_{BH}$ the black hole mass.}, we searched for X-ray sources close to the Fornax center. This search was made difficult owing to the
uncertainty with which the Fornax dSph center position is known. This is caused mainly by the large asymmetry in the galaxy surface brightness as first observed by \citet{hodge} (see also \citealt{hodge1974}). The asymmetry
 was also confirmed by \citet{stetson} who
evaluated the global galaxy centroid using a circle moving on the sky plane until the median (x, y) position of all the stars contained within
the aperture coincided with the center of the circle.
Depending on the circle radius (500 or 2000 pixels), \citet{stetson} obtained two center positions (separated by $\simeq 3'$) indicated by the green
diamonds (with labels {\it SHS98 500 pxl} and {\it SHS98 2000 pxl}) in Figure \ref{figcenters}. In the same figure,
we also give the coordinates of the centroid as obtained by \citet{hodge1974} (red diamond) and \citet{demers} (black diamond). For comparison,
the yellow diamond (on a CCD gap of the pn camera) represents the centroid as recently determined by \citet{battaglia} and the blue diamond is centered on the NED coordinates of Fornax dSph.
%%%%%%%%%%%%%%%%%%%%%%%%%%%%%%%%%%%%%%%%%%%%%%
%%%%%%%%%%%%%%%%%%%%%%%%%%%%%%%%%%%%%%%%%%%%%%
% %%%%%%%%%%%%%%%%%%%%%%%%%%%%%%%%%%%%%%%%%%%%%%
% \begin{figure*}[!t]
% \begin{center}
% \hspace{-1.08cm}
% \psfig{figure=fig1_b.eps,width=10.0cm,angle=0}
% \end{center}
% \vspace{-0.7cm}
% \caption[]{A zoomed view around the Fornax dSph center. The diamonds indicate the global centroids of the galaxy obtained by several authors (see text for details).}
% \label{figcenters}%
% \end{figure*}
% %%%%%%%%%%%%%%%%%%%%%%%%%%%%%%%%%%%%%%%%%%%%%%

Note that the centroid labeled {\it SHS98 2000 pxl}, as estimated by \citet{stetson} when considering the large structure of the galaxy,
is at a distance of $\simeq 28''$ from the centroid position according to \citet{battaglia} and is apparently very close to one 
of the X-ray sources (labeled 2)
detected in our analysis, thus allowing us to estimate the X-ray luminosity from a putative IMBH. In all the other cases, the centroids fall in a region apparently
free of sources, so that only an upper limit on the IMBH flux can be obtained. 
As a reference for this case we use {\it SHS98 500 pxl} as the position of the centroid.

In the following paragraphs we speculate about the possible existence of an IMBH in the Fornax
dSph and whether it is in the position labeled {\it SHS98 2000 pxl} or coincident with the centroid {\it SHS98 500 pxl}.

As in the case  of the G1 \citep{pooley2006} and M15 \citep{ho2003} globular clusters,
the X-ray emission from the putative IMBH at the center of the Fornax dSph may be due to Bondi accretion (\citealt{bondi1994})
onto the black hole, either from the cluster gas or from stellar winds. Thus, assuming that low-angular momentum gas
close to the compact object accretes spherically, for a black hole of mass $M_{BH}$ moving with velocity $v$ through
a gaseous medium  with hydrogen number density
$n$, the accretion rate is
\begin{equation}
\dot{M}\simeq 4\pi(GM_{BH})^2(v^2+c_s^2)^{-3/2}m_p n~,
\end{equation}
where $m_p$ and $c_s$ are the proton mass and sound speed in the medium, respectively.
The expected X-ray luminosity is then
\begin{equation}
L_X\simeq \epsilon \eta \dot{M} c^2~,
\end{equation}
which can be parametrized as
\begin{equation}
L_X\simeq \epsilon \eta 8.8\times10^{36}\left(\frac{M_{BH}}{10^3 ~{\rm M}_{\odot}}\right)^2 \left(\frac{V}{15~{\rm km~s^{-1}}}\right)^{-3}
\left(\frac{n}{0.1 ~{\rm cm^{-3}}}\right)~{\rm erg~s^{-1}~cm^{-2}}~,
\label{explum}
\end{equation}
where $V=(v^2+c_s^2)^{1/2}$, $\epsilon$ is the efficiency in converting mass to radiant energy and $\eta$ is
the fraction of the Bondi-Hoyle accretion rate onto the black hole.

The hydrogen number density of the mass feeding the black hole can be estimated by using the structural parameters of the Fornax dSph as given
in \citet{McConnachie2012}. In particular, it was found that the galaxy hosts at least $M_{HI}\simeq0.17\times 10^6$ M$_{\odot}$ of gas within
the observed half-light radius of $r_{h}\simeq710$ pc. Thus, a lower limit to the gas density can be evaluated as
\begin{equation}
n\simeq \frac{3 M_{HI}}{4\pi r_h^3 m_p}\simeq 5\times 10^{-3}~{\rm cm^{-3}}~.
\end{equation}
Assuming that $v \simeq c_s\simeq 10$ km s$^{-1}$, one has $V\simeq 14-15$ km s$^{-1}$. Thus, using the
IMBH upper limit of $M_{BH} = 3.2\times 10^4$ M$_{\odot}$ quoted above,
Eq. \ref{explum} finally gives the expected X-ray luminosity, i.e.,
\begin{equation}
L_X\simeq \epsilon \eta 4.50\times10^{38}~{\rm erg~s^{-1}}~.
\end{equation}
For an estimated distance of 0.138 Mpc to the Fornax dSph, the expected IMBH luminosity $L_X$ corresponds to an observable flux of
$F_X \simeq \epsilon \eta 1.98 \times 10^{-10}$ erg s$^{-1}$ cm$^{-2}$.

As one can see from a close inspection of Figure \ref{figcenters}, in the case when the putative IMBH position coincides 
with the centroid {\it SHS98 500 pxl}
\footnote{In the case in which the Fornax centroid position coincides with one of those evaluated 
by \citet{hodge1974}, \citet{demers}, or \citet{battaglia} (red, black, and yellow diamonds
in Figure \ref{figcenters}, respectively), we can set only an upper limit to the IMBH unabsorbed flux
of $\simeq 2.5\times 10^{-15}$ erg s$^{-1}$ cm$^{-2}$ which, at the distance of the Fornax dSph, corresponds to
a luminosity limit of $\simeq 5.7\times 10^{33}$ erg s$^{-1}$.},
there was no clear detection of X-ray counterparts in the $0.2-12$
keV band. To be conservative, the expected flux can be compared with the minimum (unabsorbed)
flux ($\simeq 2.5\times 10^{-15}$ erg s$^{-1}$ cm$^{-2}$,
see Section 2.4 for further details) detectable
in the {\it XMM}-Newton observation. Hence, one easily constrains the accretion efficiency of the IMBH to be $\epsilon \eta \leq 1.3\times 10^{-5}$.

On the contrary, a detection appears (source labeled 2) on the centroid {\it SHS98 2000 pxl} position.
In this case, using the flux values obtained by the automatic procedure described in Section 2,
we found that source 2 has an unabsorbed flux (obtained via webPIMMS) in the $0.2-12$ keV of
$\simeq 3.0\times 10^{-15}$ erg s$^{-1}$ cm$^{-2}$ which, at the Fornax dSph distance, corresponds
to an intrinsic luminosity of $L_X=7.0\times 10^{33}$ erg s$^{-1}$. Thus, in the IMBH hypothesis,
the accretion efficiency turns out to be $\epsilon \eta \simeq 1.6\times 10^{-5}$.

With a more detailed analysis, we further extracted the MOS 1, MOS 2, and pn spectra of source 2
using a circular region centered on the target coordinates and with radius
of $35''$. We then accumulated the background on a region located on the same chip and in
a position apparently free of sources. This resulted in MOS 1, MOS 2, and pn spectra with
$\simeq230$, $\simeq250$, and $\simeq480$ counts, respectively.

After grouping the data by 25 counts/bin, we imported the resulting
background-reduced spectrum within XSPEC. A model consisting of an absorbed power-law, with hydrogen column density
fixed to the average value observed towards the target (2.7$\times$10$^{20}$ cm$^{-2}$, see Section 2 for further details), resulted
in the best fit parameters $\Gamma=1.3^{+1.0}_{-0.8}$ and $N= (5.0 \pm 4.0)\times 10^{-7}$ kev$^{-1}$ cm$^{-2}$ s$^{-1}$ with
$\chi ^2_\nu =1.4$ (for 34 d.o.f.). The 0.2-12 keV band absorbed flux of source 2 is then $(0.6^{+1.7}_{-0.4})\times10^{-14}$ erg s$^{-1}$ cm$^{-2}$ (90$\%$ confidence level)
which corresponds to an unabsorbed flux of $\simeq 6.2 \times 10^{-15}$ erg s$^{-1}$ cm$^{-2}$ and an unabsorbed luminosity of
$L_X \simeq 1.4 \times 10^{34}$ erg s$^{-1}$. Assuming a spherical accretion, one finally has an
accretion efficiency of $\epsilon \eta \simeq 3\times 10^{-5}$.

Note also that for $\eta = 1$, the radiative efficiency $\epsilon$ of the putative Fornax dSph IMBH is similar to the values
found for other systems in the radiatively inefficient regime ($10^{-5}-1$, see \citealt{baganoff}, but also \citealt{fender2003}, and
\citealt{koerding2006a,koerding2006b}). The same conclusion can be reached when comparing the observed X-ray luminosity in the $0.2-12$
keV band of source 2 ($L_X\simeq 1.4\times 10^{34}$ erg s$^{-1}$) with the expected Eddington luminosity
(i.e., $L_{Edd}\simeq 1.3 \times 10^{38} \left(M_{BH}/M_{\odot}\right)$ erg s$^{-1}$)
in the IMBH hypothesis. In this case, $L_X/L_{Edd}\simeq 3\times 10^{-9}$ is obtained, thus
implying that the IMBH must be extremely radiatively inefficient (for a comparison see the M15 case discussed in
\citealt{ho2003}).
Accretion onto a black hole may be different from a simple spherical model, since the accreting gas has angular momentum. In such a case
a Keplerian disk forms and the accretion occurs due to the presence of viscous torques,
which transport the angular momentum outward from the inner to the outer regions of the disk
(see e.g. \citealt{shapiro}). In the simplified case of a thin disk structure (see \citealt{shakura}),
the total luminosity of the disk (mainly originating in its innermost parts) is
\begin{equation}
L\simeq\frac{1}{2}\frac{GM_{BH}\dot{M}}{r_I}~,
\label{diskluminosity}
\end{equation}
where $\dot{M}$ is the accretion rate and $r_I$ is the radius of the inner edge of the disk.
For a disk that extends inward to the last stable orbit ($r_I= 6GM_{BH}/c^2$ for a non-rotating black hole),
it is easy to verify that the efficiency in converting the accreting
mass to radiant energy is $\simeq 8.3\%$. When comparing the observed X-ray luminosity
in the $0.2-12$ keV band of source 2 with the value expected in the thin disk scenario
(Eq. \ref{diskluminosity}), we can estimate a mass accretion rate of $\dot{M}\simeq1.87\times 10^{14}$ g s$^{-1}$.
Assuming again the radiation efficiency to be $8.3\%$, a black hole of mass $3.2\times 10^4$ M$_{\odot}$ has an
Eddington mass accretion rate of $\dot{M}_{Edd}\simeq 5.5\times 10^{22}$ g s$^{-1}$. Thus,
the ratio $\dot{M}/\dot{M}_{Edd}\simeq4\times 10^{-9}$ implies that the IMBH in Fornax dSph (if any) accretes
very inefficiently even in the context of a Keplerian thin disk model.

The above considerations, the observed low luminosity, and the estimated power-law index
$\Gamma=1.3^{+1.0}_{-0.8}$ (marginally consistent with an index in the range 1.4-2.1, see e.g. \citealt{remilard} for a classification of black hole binaries)
allow us to depict a scenario in which the Fornax dSph IMBH may be in a quiescent state.

Recently, it has also been proposed that a relationship between black hole mass, X-ray luminosity, and radio
luminosity does exist (see e.g. \citealt{merloni} and \citealt{koerding2006a}). This {\it fundamental plane} can be
used to  test the IMBH hypothesis in globular clusters and dwarf spheroidal galaxies (\citealt{maccarone2005}).
In particular, \citet{maccarone2004} scaled the fundamental-plane relation to values appropriate for an
IMBH host in a Galactic globular cluster; by rescaling their equation to estimate the expected IMBH radio
flux at $5$ GHz, i.e.,
\begin{equation}
F_{5~{\rm GHz}}=10\left(\frac{L_X}{3\times10^{31}~{\rm cgs}}\right)^{0.6}\left(\frac{M_{BH}}{100~{\rm M_{\odot}}}\right)^{0.78}\left(\frac{10~{\rm kpc}}{d}\right)^{2}~{\rm \mu Jy}~,
\label{radio}
\end{equation}
which, for the above X-ray flux estimates, corresponds to
%
% 31710/2012. Ci siamo accorti di un errore di scrittura: nel paragrafo successivo. Poiche abbiamo gia' sottomesso
% correggiamo "in silenzio"...
%
% \begin{equation}
% F_{5~{\rm GHz}}\simeq1.3\left(\frac{\epsilon}{10^{-5}}\right)^{0.6}~{\rm mJy}~.
% \label{radio2}
% \end{equation}
% For an accretion efficiency as low as $10^{-5}$, the expected radio flux is well within the
% detection possibilities of the Australia Telescope Compact Array (ATCA) which, for an integration time of $\simeq 12$ hr, may reach
% the sensitivity of $\simeq 0.22$ mJy/beam at 5 GHz.
\begin{equation}
% % % % % % % % F_{5~{\rm GHz}}\simeq1.3\left(\frac{\epsilon}{8\times10^{-4}}\right)^{0.6}~{\rm mJy}~.
F_{5~{\rm GHz}}\simeq0.1\left(\frac{\epsilon}{10^{-5}}\right)^{0.6}~{\rm mJy}~.
\label{radio2}
\end{equation}
For an accretion efficiency as low as a few $10^{-5}$, the expected radio flux is well within the
detection possibilities of the Australia Telescope Compact Array (ATCA) which, for an integration time of $\simeq 12$ hr, may reach
an RMS sensitivity of $\simeq 10$ $\mu$Jy/beam at 5 GHz.

%--------------------------------------------------------------- TABLE
\section{Conclusions}

In this paper we re-analyzed a deep archive {\it XMM}-Newton data of the Fornax dSph galaxy with the aim of characterizing the
X-ray point-like source population. By using a restrictive analysis, we detected 107 X-ray sources. Most of them are likely
to be background objects since the number of detected sources is statistically consistent with that expected from the logN-logS relation. However,
we cannot exclude that a few of the detected objects belong to the Fornax dSph. The color-color diagram (based on the ratio between the
$0.2-2.4$ keV flux ($F_X$) and the NIR flux in J band ($F_J$) versus the J-K color) for the X-ray sources with a counterpart in the 2MASS catalog shows the presence
of a few objects (see bottom-left part of Figure \ref{figclass}) possibly of stellar nature and local origin. Furthermore, source number 61 appears to be spatially
coincident (within $\simeq 2''$) with a long-period variable star found in the catalog compiled by \citet{varstar}.

As discussed previously, among all the other dwarf spheroidal satellites of the Milky Way, a peculiar characteristic of the Fornax dSph is that
of hosting five globular clusters. We noted that two X-ray sources, detected towards the galaxy with our analysis, are coincident with the two
Fornax globular clusters GC 3 and GC 4. In particular, for GC 4 we identified one X-ray source (labeled 28 in the source list appearing in 
Table \ref{fornaxsources})
which is at a distance of $\simeq 37''$ from the globular cluster center. Consequently, considering the GC 4 structural parameters reported in
Table \ref{tableGC}, one observes that source 28 is well within the tidal radius ($\simeq 43''$) of the globular cluster and, hence, possibly associated with it (as already claimed by
\citealt{orioproc}). In the case of GC 3, we identified one X-ray source (labeled 103 at a distance of $\simeq 72''$ from the GC center) 
that might be at the outskirts of the globular cluster.

Finally, we discussed the IMBH hypothesis and found that one of the X-ray sources (labeled 2) might be associated with one
of the possible galaxy centroids identified by \citet{stetson}. In this framework, we estimated the IMBH accretion parameter 
to be $\epsilon \eta \simeq 10^{-5}$. However, since
there is a large uncertainty in the identification of the galaxy's center of gravity, the latter value can be considered as
an upper limit to the IMBH accretion parameters.
% % % % This issue could be settled by moderately deep radio and high angular resolution
% % % % X-ray  observations.

\begin{acknowledgements}
We are grateful to F. Strafella, V. Orofino, and B.M.T. Maiolo for the interesting discussions while preparing the manuscript. We also acknowledge the anonymous Referee
for his suggestions.
\end{acknowledgements}

% %%%%%%%%%%%%%%%%%%%%%%%%%%%%%%%% LONG TABLE 1
\addtocounter{table}{1}
\longtab{1}{
\footnotesize{
\begin{longtable}{|l|c|c|c|c|c|c|c|c|c|}
\caption{\label{fornaxsources}  The X-ray sources detected by our analysis towards the Fornax dSph.}\\
\hline\hline
 Src & \# & RA & Dec & Err &$HR_1$ & $HR_2$ & F$_{0.2-12 keV}^{Abs}$ & 2MASS &  USNO-B1 \\& &  (J2000) &(J2000)&(arcsec)&   & & ($\times 10^{-14}$ cgs) & (arcsec) &(arcsec) \\
\hline \endfirsthead
\caption{continued.}\\
\hline\hline
 Src & \# & RA & Dec & Err &$HR_1$ & $HR_2$ & F$_{0.2-12 keV}^{Abs}$ & 2MASS & USNO-B1 \\& &  (J2000) &(J2000)&(arcsec)&   & & ($\times 10^{-14}$ cgs) & (arcsec) & (arcsec) \\
\hline
\endhead
\hline
\endfoot
1  & E &   2 40 50.8 & -34 27 24.9 &   3.2 & -0.17 $\pm$  0.28 & -0.59 $\pm$  0.38 &  0.22 $\pm$  0.16  & & 1.5\\
2  & E &   2 39 54.5 & -34 30 41.2 &   2.6 & -0.16 $\pm$  0.33 & -0.19 $\pm$  0.33 &  0.28 $\pm$  0.22  & & \\
3  & E &   2 39 46.9 & -34 29 11.2 &   2.5 & -0.50 $\pm$  0.42 &  0.26 $\pm$  0.41 &  0.30 $\pm$  0.23  & & \\
4  & E &   2 40 47.6 & -34 20 11.3 &   2.8 & -0.15 $\pm$  0.41 & -0.62 $\pm$  0.39 &  0.31 $\pm$  0.78  & & 3.3\\
5  & E &   2 39 59.3 & -34 31  3.6 &   2.6 & -0.14 $\pm$  0.39 &  0.23 $\pm$  0.37 &  0.31 $\pm$  0.20  & & \\
6  & E &   2 39 56.1 & -34 21 16.4 &   2.8 & -0.18 $\pm$  0.37 & -0.32 $\pm$  0.38 &  0.35 $\pm$  0.29  &0.9 & 0.9\\
7 \tablefoottext{*}& E &   2 40 32.0 & -34 27 50.9 &   2.4 &  0.01 $\pm$  0.19 & -0.81 $\pm$  0.25 &  0.36 $\pm$  0.26  & & \\
8  & C &   2 39 39.4 & -34 22 58.0 &   2.4 & -0.07 $\pm$  0.40 & -0.07 $\pm$  0.36 &  0.37 $\pm$  0.31 & &  \\
9  & G &   2 39 55.6 & -34 25 53.8 &   2.7 &  0.03 $\pm$  0.37 &  0.18 $\pm$  0.35 &  0.41 $\pm$  0.22 & &  \\
10 & D &   2 39 39.1 & -34 32 21.9 &   2.3 & -0.32 $\pm$  0.27 & -0.33 $\pm$  0.29 &  0.46 $\pm$  0.40 &1.0 & 1.4\\
11 & E &   2 39 56.3 & -34 22 33.4 &   2.7 &  0.18 $\pm$  0.31 &  0.04 $\pm$  0.28 &  0.49 $\pm$  0.21 & &  \\
12 & E &   2 39 34.9 & -34 33 29.6 &   2.8 &  0.03 $\pm$  0.32 & -0.27 $\pm$  0.31 &  0.52 $\pm$  0.48 & &  \\
13 & E &   2 39 28.5 & -34 22 29.5 &   2.8 &  0.28 $\pm$  0.43 &  0.26 $\pm$  0.33 &  0.55 $\pm$  0.48 & &  \\
14  & G &   2 39 50.2 & -34 33 60.0 &   2.1 & -0.01 $\pm$  0.35 & -0.05 $\pm$  0.30 &  0.75 $\pm$  0.73 & & \\
15 & A &   2 40 20.4 & -34 24 25.5 &   2.1 & -0.10 $\pm$  0.23 & -0.12 $\pm$  0.23 &  0.75 $\pm$  0.32 & &  \\
16 & D &   2 40 21.5 & -34 23 47.1 &   2.3 & -0.05 $\pm$  0.29 & -0.05 $\pm$  0.27 &  0.83 $\pm$  0.40 & &  \\
17 & D &   2 40 38.0 & -34 36  0.3 &   2.2 & -0.09 $\pm$  0.44 & -0.01 $\pm$  0.33 &  0.84 $\pm$  0.81 & &  \\
18 & E &   2 40 19.3 & -34 19  0.2 &   2.5 & -0.06 $\pm$  0.29 & -0.19 $\pm$  0.26 &  0.84 $\pm$  0.52 & &  \\
19 \tablefoottext{*} & E &   2 40 27.0 & -34 30 24.3 &   2.4 &  0.06 $\pm$  0.24 & -0.23 $\pm$  0.24 &  0.85 $\pm$  0.39 &1.3 &1.7 \\
20 & A &   2 39 38.5 & -34 30 14.0 &   2.3 & -0.18 $\pm$  0.28 & -0.24 $\pm$  0.27 &  0.87 $\pm$  0.50 & &  \\
21 & D &   2 40 46.0 & -34 21 28.8 &   2.5 &  0.05 $\pm$  0.42 & -0.30 $\pm$  0.39 &  0.92 $\pm$  0.83 & &  \\
22 & A &   2 40 12.3 & -34 29  0.1 &   2.3 & -0.15 $\pm$  0.32 &  0.03 $\pm$  0.29 &  0.93 $\pm$  0.47 & &  \\
23 & D &   2 39 17.7 & -34 16 53.9 &   2.3 & -0.33 $\pm$  0.39 &  0.14 $\pm$  0.30 &  0.95 $\pm$  0.83 & &  \\
24  & G &   2 38 58.1 & -34 29 22.0 &   2.2 & -0.28 $\pm$  0.47 &  0.13 $\pm$  0.36 &  0.95 $\pm$  1.15 & &  \\
25 & E &   2 39 54.5 & -34 22  6.7 &   2.5 &  0.10 $\pm$  0.29 &  0.02 $\pm$  0.24 &  0.96 $\pm$  0.36 & &  \\
26 & E &   2 39  4.1 & -34 37 22.7 &   2.5 & -0.31 $\pm$  0.26 & -0.24 $\pm$  0.25 &  0.98 $\pm$  0.75 & &  \\
27 & C &   2 39 32.4 & -34 29 51.8 &   2.2 & -0.11 $\pm$  0.25 & -0.37 $\pm$  0.29 &  0.99 $\pm$  0.50 & &  \\
28 & A &   2 40  8.8 & -34 31 36.4 &   2.2 & -0.16 $\pm$  0.22 & -0.06 $\pm$  0.21 &  1.01 $\pm$  0.40 & &  \\
29 & E &   2 39 50.7 & -34 40 36.3 &   2.3 &  0.05 $\pm$  0.18 & -0.93 $\pm$  0.21 &  1.03 $\pm$  0.92 & &  \\
30  & G &   2 40 25.9 & -34 19 58.9 &   2.4 &  0.07 $\pm$  0.47 &  0.43 $\pm$  0.38 &  1.04 $\pm$  0.70 & & \\
31 & C &   2 40 18.9 & -34 27 23.3 &   2.4 &  0.16 $\pm$  0.30 &  0.10 $\pm$  0.22 &  1.10 $\pm$  0.44 & &  \\
32 & E &   2 39 33.2 & -34 26 41.4 &   2.7 &  0.42 $\pm$  0.36 &  0.10 $\pm$  0.28 &  1.10 $\pm$  0.44 & &  \\
33 & D &   2 39 20.2 & -34 27 16.0 &   2.2 & -0.08 $\pm$  0.20 & -0.25 $\pm$  0.22 &  1.12 $\pm$  0.52 & &  \\
34 & A &   2 39 41.4 & -34 24 45.5 &   2.2 & -0.12 $\pm$  0.36 &  0.09 $\pm$  0.30 &  1.22 $\pm$  0.67 & &  \\
35 & D &   2 40 30.4 & -34 32 46.5 &   2.5 &  0.09 $\pm$  0.32 & -0.08 $\pm$  0.26 &  1.26 $\pm$  0.70 & &  \\
36 & E &   2 40 35.1 & -34 36 28.8 &   2.5 &  0.16 $\pm$  0.36 & -0.45 $\pm$  0.35 &  1.26 $\pm$  1.02 & &  \\
37 & A &   2 40 11.2 & -34 30 51.4 &   2.2 & -0.01 $\pm$  0.24 &  0.02 $\pm$  0.22 &  1.27 $\pm$  0.45 & &  \\
38 & A &   2 39 37.7 & -34 36 56.8 &   2.3 & -0.21 $\pm$  0.27 & -0.08 $\pm$  0.23 &  1.33 $\pm$  0.63 & &  \\
39 & A &   2 40  4.5 & -34 28 55.1 &   2.3 & -0.27 $\pm$  0.14 & -0.17 $\pm$  0.15 &  1.41 $\pm$  0.37 & &  \\
40 & C &   2 39 18.8 & -34 26 43.4 &   2.5 &  0.08 $\pm$  0.28 & -0.11 $\pm$  0.25 &  1.45 $\pm$  0.72 &2.5 &2.0 \\
41 \tablefoottext{*}& A &   2 39 25.5 & -34 17 35.0 &   2.1 & -0.05 $\pm$  0.18 & -0.73 $\pm$  0.22 &  1.51 $\pm$  1.11 & 2.4 &  2.0\\ % riga nuova inserita e rinominata 41
42 & A &   2 40 33.4 & -34 25 12.6 &   2.2 &  0.03 $\pm$  0.26 &  0.00 $\pm$  0.21 &  1.62 $\pm$  0.65 & &  \\ % riga 41 della tabella vecchia rinominata in 42
43 & E &   2 40  3.4 & -34 21  5.3 &   2.1 & -0.16 $\pm$  0.13 & -0.15 $\pm$  0.12 &  1.63 $\pm$  0.41 & &  \\ % riga 42 della tabella vecchia rinominata in 43
%43 \tablefoottext{*}& D &   2 39 25.4 & -34 17 35.5 &   2.1 & -0.06 $\pm$  0.15 & -0.71 $\pm$  0.19 &  1.65 $\pm$  0.96 & &2.0 & &1.6 \\ % riga 43 della tabella vechhia eliminata
44 & E &   2 40  1.5 & -34 17 51.0 &   2.3 &  0.01 $\pm$  0.21 & -0.07 $\pm$  0.18 &  1.70 $\pm$  0.67 & &  \\
45 & E &   2 40 32.4 & -34 18  1.9 &   2.6 &  0.36 $\pm$  0.39 &  0.25 $\pm$  0.32 &  1.73 $\pm$  0.85 & &  \\
46 & A &   2 40 19.6 & -34 32 52.7 &   2.1 & -0.16 $\pm$  0.17 & -0.19 $\pm$  0.17 &  1.76 $\pm$  0.70 &0.1 &1.1 \\
47 & E &   2 39 28.5 & -34 33 20.0 &   4.4 & -0.03 $\pm$  0.31 & -0.51 $\pm$  0.33 &  1.79 $\pm$  1.50 & &  \\
48 & A &   2 40 25.9 & -34 33 28.2 &   2.1 & -0.14 $\pm$  0.12 & -0.55 $\pm$  0.15 &  1.86 $\pm$  0.71 &1.4 &1.4 \\
49 & D &   2 38 55.4 & -34 30  0.3 &   2.6 & -0.07 $\pm$  0.28 & -0.16 $\pm$  0.25 &  1.89 $\pm$  1.18 & &  \\
50 & D &   2 39 35.0 & -34 32  7.8 &   2.2 & -0.06 $\pm$  0.22 & -0.02 $\pm$  0.18 &  1.95 $\pm$  0.87 & &  \\
51 & D &   2 39 37.7 & -34 19 58.1 &   2.4 &  0.61 $\pm$  0.31 &  0.13 $\pm$  0.22 &  1.99 $\pm$  0.80 & &  \\
52 & G &   2 39 44.5 & -34 18 27.6 &   2.2 & -0.32 $\pm$  0.18 & -0.02 $\pm$  0.17 &  2.00 $\pm$  0.71 & &  \\
53 & F &   2 39 25.5 & -34 21 21.3 &   2.3 & -0.08 $\pm$  0.35 & -0.15 $\pm$  0.34 &  2.01 $\pm$  1.25 & & 1.9\\
54 & E &   2 40 19.0 & -34 34 43.8 &   2.4 &  0.26 $\pm$  0.28 &  0.00 $\pm$  0.17 &  2.02 $\pm$  0.80 & & \\
55 & D &   2 39  5.1 & -34 18 52.9 &   2.3 &  0.13 $\pm$  0.33 & -0.68 $\pm$  0.37 &  2.12 $\pm$  1.82  & &0.7 \\
56 & D &   2 40 37.5 & -34 24  3.6 &   2.3 &  0.05 $\pm$  0.21 & -0.09 $\pm$  0.18 &  2.12 $\pm$  0.74  & & \\
57  & G &   2 39 38.6 & -34 20 21.6 &   2.1 &  0.27 $\pm$  0.39 & -0.26 $\pm$  0.35 &  2.16 $\pm$  1.15 & & 2.2\\
58 & E &   2 39 13.4 & -34 22 38.8 &   2.6 & -0.03 $\pm$  0.30 &  0.10 $\pm$  0.21 &  2.21 $\pm$  0.97  & & \\
59 & C &   2 39 43.7 & -34 30 40.5 &   2.3 &  0.48 $\pm$  0.29 &  0.26 $\pm$  0.20 &  2.25 $\pm$  0.65  & & \\
60 & E &   2 38 57.2 & -34 26 30.0 &   2.7 &  0.20 $\pm$  0.38 &  0.14 $\pm$  0.25 &  2.26 $\pm$  1.40  & & \\
61\tablefoottext{V} & A &   2 39 41.4 & -34 33 38.0 &   2.1 & -0.10 $\pm$  0.16 & -0.20 $\pm$  0.17 &  2.28 $\pm$  0.71  & & \\   % 2.0 arcsecondi di distanza dalla sorgente var*
62  & G &   2 40 47.3 & -34 29  7.7 &   2.0 &  0.64 $\pm$  0.30 &  0.17 $\pm$  0.18 &  2.31 $\pm$  1.28 & & \\
63 & A &   2 40 45.6 & -34 33 50.1 &   2.1 & -0.16 $\pm$  0.26 &  0.00 $\pm$  0.19 &  2.36 $\pm$  1.25 & &  \\
64 & E &   2 39 33.0 & -34 21 11.1 &   2.6 &  0.46 $\pm$  0.27 &  0.05 $\pm$  0.19 &  2.46 $\pm$  0.74 & &  \\
65 & A &   2 40 43.4 & -34 25 16.7 &   2.2 & -0.10 $\pm$  0.20 & -0.19 $\pm$  0.19 &  2.48 $\pm$  1.02 & & 2.4\\
66 & A &   2 39 58.0 & -34 28 17.0 &   2.1 &  0.21 $\pm$  0.23 &  0.36 $\pm$  0.16 &  2.52 $\pm$  0.57  & & \\
67 & B &   2 40 29.4 & -34 25 10.3 &   2.1 &  0.12 $\pm$  0.30 &  0.37 $\pm$  0.22 &  2.53 $\pm$  0.91  & & \\
68 & E &   2 38 58.1 & -34 34  2.5 &   3.0 &  0.10 $\pm$  0.38 & -0.13 $\pm$  0.26 &  2.64 $\pm$  1.85  & & \\
69 & B &   2 40 58.6 & -34 23 56.6 &   2.3 & -0.12 $\pm$  0.19 & -0.31 $\pm$  0.22 &  2.67 $\pm$  1.05 & & 0.6\\
70 & D &   2 40 52.0 & -34 29 45.7 &   2.1 &  0.30 $\pm$  0.37 & -0.06 $\pm$  0.24 &  2.68 $\pm$  1.26 & & \\
71  & G &   2 39 58.5 & -34 13 49.7 &   2.2 &  0.07 $\pm$  0.32 &  0.13 $\pm$  0.24 &  2.76 $\pm$  1.61 & & \\
72 & B &   2 40 58.0 & -34 30  9.5 &   2.1 &  0.03 $\pm$  0.29 & -0.09 $\pm$  0.25 &  2.99 $\pm$  1.62  & & \\
73 & D &   2 39 48.5 & -34 17 16.6 &   2.3 &  0.24 $\pm$  0.30 &  0.36 $\pm$  0.22 &  3.01 $\pm$  1.01  & & 2.0\\
74 & D &   2 39 40.4 & -34 20  2.5 &   2.2 & -0.05 $\pm$  0.18 &  0.17 $\pm$  0.15 &  3.33 $\pm$  0.87  & & \\
75 & A &   2 39 52.6 & -34 34 26.2 &   2.3 & -0.11 $\pm$  0.23 &  0.03 $\pm$  0.19 &  3.58 $\pm$  1.19  &1.1&0.7 \\
76  & G &   2 41  2.6 & -34 28 48.8 &   2.1 &  0.01 $\pm$  0.41 &  0.45 $\pm$  0.29 &  3.81 $\pm$  2.00 & & 1.4\\
77 & B &   2 39 13.2 & -34 17 28.9 &   2.2 & -0.05 $\pm$  0.34 &  0.18 $\pm$  0.28 &  3.87 $\pm$  2.12 & & \\
78 & G &   2 41  3.8 & -34 29  9.4 &   2.1 & -0.04 $\pm$  0.32 & -0.05 $\pm$  0.25 &  3.96 $\pm$  2.09  & & \\
79 & A &   2 40  8.6 & -34 23 21.6 &   2.1 & -0.10 $\pm$  0.11 & -0.03 $\pm$  0.10 &  4.09 $\pm$  0.64 & & \\
80 & A &   2 39 43.2 & -34 33 37.5 &   2.1 &  0.21 $\pm$  0.22 &  0.33 $\pm$  0.16 &  4.17 $\pm$  1.05 & & \\
81 & D &   2 39 14.4 & -34 34  0.1 &   2.3 &  0.80 $\pm$  0.23 &  0.01 $\pm$  0.14 &  4.39 $\pm$  1.62  & & \\
82 \tablefoottext{*} & A &   2 39 34.0 & -34 21 51.1 &   2.0 & -0.24 $\pm$  0.09 & -0.27 $\pm$  0.09 &  4.41 $\pm$  0.66 &0.4 &0.8 \\
83 & A &   2 39 21.4 & -34 34 26.7 &   2.2 &  0.04 $\pm$  0.25 &  0.20 $\pm$  0.20 &  4.45 $\pm$  1.68 & &  \\
84 & A &   2 40 33.7 & -34 27  3.7 &   2.1 & -0.22 $\pm$  0.07 & -0.44 $\pm$  0.09 &  4.71 $\pm$  0.63 &1.7 &1.7 \\
%85 & D &   2 40 32.2 & -34 36 38.3 &   2.1 &  0.06 $\pm$  0.20 &  0.24 $\pm$  0.15 &  5.05 $\pm$  1.56 & & & & \\ riga 85 della tabella vecchia rimpiazzata dalla 85 nuova messa di seguito
85 & A &   2 40 32.1 & -34 36 39.2 &   2.1 &  0.03 $\pm$  0.24 &  0.29 $\pm$  0.18 &  4.96 $\pm$  1.77 & & \\ % riga 85 nuova che rimpiazza la 85 della tabella vecchia
86 & A &   2 39 43.9 & -34 26 15.9 &   2.2 &  0.79 $\pm$  0.21 &  0.00 $\pm$  0.14 &  5.12 $\pm$  0.97 &0.7&1.2 \\
87 & C &   2 39 49.2 & -34 32 35.2 &   2.1 &  0.11 $\pm$  0.16 &  0.06 $\pm$  0.12 &  5.39 $\pm$  1.16 & &  \\
88 & D &   2 39  3.6 & -34 24 40.2 &   2.4 &  0.25 $\pm$  0.32 &  0.25 $\pm$  0.23 &  5.91 $\pm$  2.53 & &  \\
89 & E &   2 38 56.5 & -34 32 14.3 &   2.5 &  0.24 $\pm$  0.25 &  0.15 $\pm$  0.17 &  6.09 $\pm$  1.96 & &  \\
90 & E &   2 39  3.5 & -34 36 39.1 &   2.1 & -0.04 $\pm$  0.14 & -0.18 $\pm$  0.12 &  6.28 $\pm$  1.90 & & 1.7\\
91 & F &   2 39 17.9 & -34 31 59.8 &   2.2 &  0.47 $\pm$  0.31 &  0.04 $\pm$  0.20 &  6.44 $\pm$  2.18 & & \\
92 & A &   2 39 13.0 & -34 17  8.3 &   2.2 & -0.12 $\pm$  0.18 & -0.09 $\pm$  0.16 &  6.80 $\pm$  2.37 & & 0.7\\
93 & E &   2 39 24.7 & -34 32 33.2 &   3.9 & -0.19 $\pm$  0.20 & -0.22 $\pm$  0.15 &  6.87 $\pm$  3.60 & & \\
94 & G &   2 40 10.0 & -34 12 30.6 &   2.2 &  0.11 $\pm$  0.26 &  0.30 $\pm$  0.19 &  7.70 $\pm$  2.68 & & \\
95 & A &   2 39  8.1 & -34 18 20.1 &   2.1 & -0.11 $\pm$  0.17 &  0.16 $\pm$  0.13 &  8.06 $\pm$  2.27 & & \\
96  & G &   2 39 43.3 & -34 14 21.3 &   2.2 &  0.13 $\pm$  0.22 &  0.03 $\pm$  0.16 &  8.28 $\pm$  2.55& & \\
97 & A &   2 40  8.1 & -34 34 19.7 &   2.0 & -0.13 $\pm$  0.08 & -0.22 $\pm$  0.08 &  8.69 $\pm$  1.26 & &2.2 \\
98 & C &   2 39 23.4 & -34 33 15.8 &   2.1 &  0.49 $\pm$  0.16 &  0.19 $\pm$  0.11 &  8.87 $\pm$  1.72 & & \\
99 & A &   2 39  3.4 & -34 31 55.4 &   2.2 &  0.40 $\pm$  0.22 &  0.25 $\pm$  0.14 &  8.97 $\pm$  2.49 & & 2.0\\
100 & G &   2 40 14.6 & -34 12 59.9 &   2.2 & -0.17 $\pm$  0.12 & -0.26 $\pm$  0.13 & 10.04 $\pm$  2.98 & & \\
101 & A &   2 39 12.4 & -34 18 16.9 &   2.1 & -0.06 $\pm$  0.13 &  0.12 $\pm$  0.10 & 10.05 $\pm$  2.19 & & \\
102 & A &   2 39 15.3 & -34 35 28.3 &   2.0 & -0.05 $\pm$  0.12 & -0.06 $\pm$  0.10 & 10.70 $\pm$  2.34 &2.0 &1.6 \\
103 & G &   2 39 49.2 & -34 14 19.2 &   2.1 &  0.04 $\pm$  0.13 & -0.16 $\pm$  0.12 & 11.69 $\pm$  2.59 & & \\
104 & D &   2 39 49.0 & -34 19 58.8 &   2.0 & -0.10 $\pm$  0.05 & -0.27 $\pm$  0.05 & 12.58 $\pm$  1.01 & & 1.7\\
105 & A &   2 39 32.8 & -34 25 25.1 &   2.0 & -0.10 $\pm$  0.05 & -0.18 $\pm$  0.06 & 13.14 $\pm$  1.05 & &1.2 \\
106 & E &   2 40 39.0 & -34 38 58.6 &   2.0 &  0.00 $\pm$  0.05 & -0.20 $\pm$  0.06 & 23.52 $\pm$  2.34 & & 2.4\\
107 \tablefoottext{*,M} & A &   2 40 19.0 & -34 37 19.9 &   2.0 & -0.05 $\pm$  0.07 & -0.08 $\pm$  0.07 & 24.79 $\pm$  3.06 &0.5 &1.0 \\ % 1.0 arcosecondi di distanza da una sorgente 2MASX

\hline
\hline
\end{longtable}
\tablefoot{
\tablefoottext{*}{An asterix close to the X-ray source label means that
the source is possibly associated (within 1 $\sigma$) with a counterpart found in the position and proper motions extended catalog (PPMX, see \citealt{roeser2008}), thus implying a possible foreground source.}
\tablefoottext{V}{When correlating the X-ray sources with the available catalogs, we found that source number 61 is possibly associated, within 2.0$''$,
to one (J023941.4-343340) belonging to a catalog of variable stars \citep{varstar} (see text for details).}
\tablefoottext{M}{The X-ray source labeled as 107 also correlates (within 1.0$''$) with a source in the 2MASX \citep{2mass} and is possibly associated to a background AGN (QJ0240-3437, see also \citealt{mendez2011}
where this quasar has been considered as a reference object for measuring the proper motion of sources belonging to the Fornax dSph). As a consequence, source 107 may have been
erroneously reported in the PPMX catalog (source labeled as 024019.0-343719).}
}
}
}% End \longtab

\end{document}